\documentclass[submission, Phys]{SciPost}
\usepackage{amssymb, amsmath}

\begin{document}

\begin{center}{\Large \textbf{
Quantum metric and localization in a quasicrystal
}}\end{center}

\begin{center}
Q. Marsal\textsuperscript{1*},
P. Holmvall\textsuperscript{1},
and A. M. Black-Schaffer\textsuperscript{1}
\end{center}

\begin{center}
{\bf 1} Department of Physics and Astronomy, Uppsala University, Box 524, 
SE-751 20 Uppsala, Sweden
\\
* quentin.marsal@physics.uu.se
\end{center}

\begin{center}
\today
\end{center}

\vspace{10pt}
\noindent\rule{\textwidth}{1pt}
\tableofcontents\thispagestyle{fancy}
\noindent\rule{\textwidth}{1pt}
\vspace{10pt}

\section*{Abstract}
{\bf
We use the quantum metric to understand the properties of quasicrystals, represented by the one-dimensional (1D) Fibonacci chain. We show that the quantum metric can relate the localization properties of the eigenstates to the self-similarity of both the chain and its energy spectrum. In particular, the quantum metric incorporates information about distances between the local symmetry centers of each eigenstate, making it much more sensitive to the localization properties of quasicrystals than other measures of localization, such as the inverse participation ratio. 
Importantly, we further find that a complete description of localization requires us to, in addition, introduce a new phasonic component to the quantum metric, along with a similarly mixed phason-position Chern number. Using this, we show that the sum of both position and phasonic components of the quantum metric is lower-bounded by the gap labels associated with each energy gap of the Fibonacci chain, which stem from the Chern number. This establishes a direct link through the quantum geometry between spatial localization and fractal energy spectrum of quasicrystals. 
Taken together, quantum geometry provides a unifying, yet accessible, understanding of quasicrystals, rooted in their self-similarity and with intriguing consequences also for many-body physics.
}

\section{Introduction}
Quasicrystalline materials have since their experimental discovery in 1984~\cite{Shechtman1984} been subject to growing interest thanks to their highly peculiar electronic properties. 
Quasicrystals are long-range ordered materials like crystals, but without translation symmetry.
Instead, the long-range order in quasicrystals stems from self-similarity of their atomic structure, visible in their diffraction pattern~\cite{Shechtman1984}.
Consequently, the emergent electronic properties of quasicrystals are closely related to their unusual atomic structure.
Recent studies have in particular focused on the localization properties of the electronic states, demonstrating their criticality~\cite{mace2017b} and topological properties~\cite{kraus2012b,varjas2019, rai2021,else2021,fan-huang2022}.
There is also growing interest in how these localization properties influence interacting~\cite{mace2019,chiaracane2021,Bonsel2026} and ordered states such as superconductivity~\cite{sakai2017,kamiya2018,zhang2022,wang2024,tokumoto2024} or its proximity effect~\cite{rai2019,rai2020} and Josephson effect~\cite{sandberg2024}, topological superconductivity~\cite{fulga2016,ghadimi2021,kobialka2024,hori2024}, antiferromagnetism~\cite{tamura2025}, and quantum information processing~\cite{Gosh2025}.
Recently, quasiperiodicity has also been found in moir{\'e} structures~\cite{lubin2012,joon-ahn2018,mahmood2021,uri2023}, further broadening the importance of quasicrystals.

Among the most studied quasicrystals is the Fibonacci chain~\cite{jagannathan21}. 
It is a chain of sites (atoms) coupled by two types of hoppings that alternate following a regular, but non-crystalline pattern.
Since it is one-dimensional (1D), it is simple to study and both numerical and analytical results of the electronic spectrum are available~\cite{kohmoto1983,ostlund1983, sire89,piechon95}.
Yet, it remains similar to more elaborate quasicrystals, as it exhibits most of the general properties of quasicrystals~\cite{ moustaj23,reisner2023,wang2024}, including the effect of hidden dimensions~\cite{kraus12,madsen13,jagannathan2025}. It has therefore come to serve as an effective model to describe the essential physics of also 2D and 3D quasicrystals~\cite{jagannathan21,moustaj23,wang2024}. Moreover, the Fibonacci chain has been experimentally realized and studied in a wide range of physical systems~\cite{negro2003,steurer2007,kraus2012b,verbin2013,tanese2014,verbin2015,baboux2017,zilberberg2018,lisiecki2019,goblot2020,reisner2023,franca2024,Ghosh2026}.

Stemming directly from the quasicrystalline nature of the Fibonacci chain, the localization properties of its electronic eigenstates are key for understanding its physics~\cite{yao2019}. 
However, so far studies have been mainly restricted to using the inverse participation ratio (IPR)~\cite{lahini2009} or multifractal exponents~\cite{mace2016, ahmed2022, reisner2023}, which only give a partial description of localization and also do not provide any direct relationship to the energy spectrum. 
Indeed, these indicators only take into account the number of occupied sites for a given state and thus do not capture correlations of the localization of the electronic state to the local atomic structure of the chain, known to produce self-similarity and multifractal states~\cite{mace2016,reisner2023}.

In this work, we establish the quantum metric as an indispensable and natural tool for providing extensive understanding of quasicrystals, capturing both their intricate localization properties and their relation to the energy spectrum. The quantum metric measures distances between eigenstates based on a parameter space, usually momentum space~\cite{provost1980}. 
It has also been proven to have a simple interpretation in real space, where it gives the localization of Wannier orbitals~\cite{marzari1997}. Building on the quantum metric efficiently probing localization, recent work has shown that disorder may enhance the quantum metric~\cite{marsal2024} when the impurity states are critically localized. 
In addition, the quantum metric has been found to be important for interacting properties of narrow or flat bands, for example by enabling superconducting phase coherence~\cite{peotta2015,tian2023} and the fractional quantum Hall effect~\cite{bergholtz2013, ledwith2020, varjas2022}.
Here, we are able to establish a direct correspondence between the structural properties of the chain and the localization of its electronic states, by showing how the quantum metric incorporates information about distances between local symmetry centers of each eigenstate and how these are inherently linked to the self-similarity of the Fibonacci chain.
This makes the quantum metric a natural tool for understanding localization in general quasicrystals, also beyond the 1D Fibonacci chain, in contrast to the IPR and other related tools traditionally used, which primarily focus only on the number of occupied sites \cite{lahini2009,mace2016, reisner2023}.
Moreover, with the quantum metric being a numerically easily accessible observable, with rapidly emerging experimental signatures \cite{neupert2013, ozawa2019, kang2025, verma2026}, it offers direct access to electronic localization of quasicrystals. 

Importantly, we need to introduce a novel formulation for the quantum metric that, in addition, takes into account the different phason realizations, encoding different incarnations of the chain, as an independent degree of freedom. We show that the full quantum metric, defined as the sum of the positional and phasonic quantum metric contributions, inherits properties from those of a 2D crystal. Importantly, we are able to provide a lower bound of the full quantum metric, set by the Chern number of the corresponding 2D crystal. Since the same Chern number labels the gaps in the energy spectrum of the quasicrystal \cite{kraus12}, we can thereby explicitly relate the full quantum metric to the self-similarity of the energy spectrum in quasicrystals. Thus, the quantum geometry provides a unified tool to understand both localization and energy spectrum in quasicrystals.
This also allows us to determine analytically the range of gap labels over which the effect of the quasicrystalline structure on the localization properties dominates over finite size effects.
For weak quasicrystal modulations we additionally establish numerically that the lower bound  constrains the positional quantum metric.
Our results moreover establish that quasicrystals are materials where the quantum geometry can be tuned over a wide range. This opens for many-body phases in quasicrystals dependent on the quantum metric, such as flat-band superconductivity~\cite{peotta2015,tian2023} and fractional quantum Hall effect in Chern insulators~\cite{bergholtz2013, ledwith2020, varjas2022}.

\section{Results}
We start by considering the quantum geometry of a Fibonacci chain. To do so, we introduce a new definition of a mixed position-phason Chern number and quantum metric in Section~\ref{sec:geoQ}. With this framework we can then analytically extract a minimal bound on the quantum metric in Section~\ref{sec:minbound}, which naturally and explicitly links the localization properties set by the quantum metric to the fractal gap structure of the energy spectrum. Further, this also allows us distinguish two regimes where the electron localization in finite approximant of the Fibonacci chain is governed either by the quasicrystalline structure of the chain or its finite size. In chains with small quasicrystal modulations we show numerically that the bound is also  applicable for the positional quantum metric. Finally, in Section~\ref{sec:renorm} we relate the dependence of the positional quantum metric on the filling fraction to the local structure and symmetries of the chain based on a renormalization scheme previously developed to understand quasicrystals. This allows us to understand why the quantum metric is a natural tool to quantify localization in quasicrystals.

\subsection{Quantum geometry of the Fibonacci chain}\label{sec:geoQ}
To start we note that the Fibonacci chain is most generally described through both a position coordinate and the so-called phason parameter, or angle, which describes different incarnations of the Fibonacci sequence \cite{kraus12,madsen13,jagannathan2025} (see App.~\ref{app:fibo} for details). Thus, the Fibonacci chain can be mapped on a 2D parameter space such that the Hamiltonian of any finite approximant of the chain reads
\begin{equation}
    \mathcal{H} = \int \mathrm{d}\phi\sum_{n=0}^{N-1} -t_{n}(\phi)c_n^\dagger(\phi)c_{n+1}(\phi) + {\rm H.c.},\label{eq:Hphi2}
\end{equation}
where $\phi$ labels the different phason modes and $n$ labels the sites and associated bonds in the chain, $N$ being the total number of sites that the approximant contains.
The dependence of the creation operators $c^{(\dagger)}_n(\phi)$ and nearest neighbor hopping $t_n(\phi)$ on $\phi$ in Eq.~\eqref{eq:Hphi2} indicates that $\phi$ is to be considered as an independent degree of freedom of the system.
Here $t_n(\phi)$ is given by~\cite{jagannathan21}
\begin{equation}
        t_n(\phi) = \frac{t_L+t_S}{2}\left(1+\delta\ \mathrm{sgn}\left[\cos\left(\frac{2n\pi q}{p+q}+\phi\right) -\cos\left(\frac{\pi q}{p+q}\right)\right]\right)\label{eq:ph_dep},
\end{equation}
with 
\begin{equation}
    \delta = \frac{t_S-t_L}{t_L+t_S}
\end{equation} 
quantifying the modulation strength of the chain, and $p$, $q$ the numbers of short and long bonds in the approximant, such that $\frac{p}{q} = \frac{2}{1+\sqrt{5}}$, the inverse golden mean.
Plotting the energy spectrum of Eq.~\eqref{eq:Hphi2} as a function of $\phi$, we find a multitude of gaps with edge states, whose winding number $\nu$ through the gap can be used as a gap label (see App.~\ref{app:fibo}).

The Hamiltonian in Eq.~\eqref{eq:Hphi2} effectively describes a 2D system, mixing a position degree of freedom $x$ and a momentum-like one $\phi$. 
Thanks to this 2D nature, we can introduce its Berry curvature.
Previously established formulations of the Berry curvature use either momentum~\cite{berry1984} or position~\cite{bianco2011} degrees of freedom, with  the former also used previously for quasicrystals to highlight similarities to a 2D Chern insulator \cite{kraus12, madsen13}.
Here, we merge these two diametrical opposite approaches into a new formulation of the Chern number, the integral of the Berry curvature. This is a more useful formulation for quasicystals as that mixes natural coordinates, position space along one dimension with phason-like momentum space along the other:
\begin{equation}
    C = \frac{1}{2i\pi N}\int_0^{2\pi}\mathrm{d}\phi\mathrm{Tr}_{\rm bulk}[P^{(\phi)}xQ^{(\phi)}\partial_\phi P^{(\phi)}
    -\partial_\phi P^{(\phi)}Q^{(\phi)}xP^{(\phi)}],
    \label{eq:MixedC}
\end{equation}
where $N$ is the number of sites in the chain and $P^{(\phi)}$, $Q^{(\phi)}$ are projectors on the occupied and unoccupied states, respectively, of the phason mode $\phi$. To capture bulk properties, we set the bulk sites over which the trace is computed when using open boundary conditions (OBCs) as (arbitrarily) the central third of the chain.

While the Berry curvature makes up the imaginary part of the quantum geometric tensor we, in addition, have to formulate a new two-component quantum metric, i.e.~the real part of the quantum geometric tensor, to incorporate both position space and phason degrees of freedom,
\begin{equation}
    \Omega = \Omega_x+\Omega_\phi,\label{eq:OxOphi}
\end{equation}
which read
\begin{equation}
    \Omega_x = \frac{1}{2\pi N}\int_0^{2\pi}\mathrm{d} \phi \mathrm{Tr}[P^{(\phi)}x Q^{(\phi)}xP^{(\phi)}]\label{eq:Ox},
\end{equation}
and
\begin{equation}
    \Omega_\phi = \frac{1}{2\pi N}\int_0^{2\pi}\mathrm{d} \phi\mathrm{Tr}\left[\partial_\phi P^{(\phi)}Q^{(\phi)}\partial_\phi P^{(\phi)}\right]\label{eq:Ophi}.
\end{equation}
Note that we could in principle also introduce the real part of the mixed component $\Omega_{x\phi}$, similar to Eq.~\eqref{eq:MixedC}, but we do not consider this component here, as only the trace expressed in Eq.~\eqref{eq:OxOphi} has a simple interpretation in terms of localization of the Wannier centers~\cite{marzari1997} and is therefore advantageous for quantifying localization (See App.~\ref{app:metric}).
Also note that, contrary to the Chern number, the components of the quantum metric can be traced over the full system also when using OBCs. This is because the edge sites do not bring a specific contribution to the quantum metric, whereas bulk and edge sites have opposite contributions to the Chern number in Eq.~\eqref{eq:MixedC}, with only the bulk sites being quantized to the Chern number~\cite{bianco2011}.

We further note that Eq.~\eqref{eq:MixedC} and Eq.~\eqref{eq:OxOphi} are defined for a continuous phason parameter, whereas the phason modes in fact remain discrete due to the sign function in Eq.~\eqref{eq:ph_dep}.
In principle, the derivatives with respect to $\phi$ are thus ill-defined for the Fibonacci chain, as the hoppings suddenly switch between $t_L$ and $t_S$ for two different phason modes.
This results in $\Omega_\phi$ diverging in the thermodynamic limit. 
Remarkably, the Chern number Eq.~\eqref{eq:MixedC} in a given gap does not diverge in the thermodynamic limit, due to the fact that it contains only one $\phi$-derivative of a sign function, which can be described in terms of Dirac-delta distributions and are in turn integrable. 
The divergence of $\Omega_\phi$ can equivalently be seen in the position space picture by decomposing Eq.~\eqref{eq:ph_dep} into a Fourier series (see App.~\ref{app:FT} for more details)
\begin{equation}
    t_n(\phi) = \sum_{l = -\infty}^{l=+\infty} \tilde{t}_n^le^{-il\phi},\label{eq:ph_depFT}
\end{equation}
where 
\begin{equation}
    \tilde{t}_n^0 = \frac{t_L+t_S}{2}\left(1+\delta\left(2\frac{q}{p+q}-1\right)\right),
\end{equation}
and for $l\neq 0$
\begin{equation}
    \tilde{t}_n^l = \frac{t_S-t_L}{2}\frac{2}{\pi l}\sin\left(\frac{\pi l q}{p+q}\right)e^{2i\pi ln\frac{q}{p+q}}.
\end{equation}
In Eq.~\eqref{eq:ph_depFT}, the term $\tilde{t}_n^l$ corresponds to a hopping from site $(n,m)$ to site $(n+1,m+l)$, where $m$, hereafter referred to as the conjugate position, labels the `position' along the phasonic direction, {\it i.e.}~the position degree of freedom conjugate to the phason parameter. 
The sharp switch from $t_L$ to $t_S$ and vice-versa as a function of $\phi$ in Eq.~\eqref{eq:ph_dep} results in a slow decay of the harmonics as a function of $l$, $|\tilde{t}_n^l|\underset{l\to\infty}{\sim} \frac{1}{l}$.
These high harmonics correspond to long-range hoppings along the position conjugate to the phasonic direction, thus resulting in fully delocalized electronic states, which generate a diverging quantum metric.

To avoid this divergence and keep a smooth variation of the Hamiltonian as a function of $\phi$, we introduce a cutoff $|l|<L$ in the series in Eq.~\eqref{eq:ph_depFT}. 
In the phasonic space, this cutoff smoothens the change between two phason modes while, in the conjugate position space picture, hoppings with range greater than $L$ become forbidden.
In principle, this cutoff influences both the spectrum and the eigenstates of the system. 
For example, gaps with label $|\nu|>L$ do not open in the smoothened version of the Fibonacci chain.
A similar effect is also known to appear in 2D Chern insulators~\cite{sticlet2013}: long-range hoppings are necessary to obtain high Chern number insulators. 
Thus, in order to keep the gap structure of the spectrum of the chain, we choose the cutoff $L$ to be the size of the approximant $N$, so that it does not close any gaps. 

\subsection{Minimal bound for the quantum metric}\label{sec:minbound}

Having introduced the quantum geometry of the Fibonacci chain in position-phason space, we next discuss how the Chern number and quantum metric are inherently linked to the gap structure of the energy spectrum and to the gap labels, and derive a number of important consequences and bounds.
First of all, the positivity of the quantum geometric tensor~\cite{peotta2015, yu2025} mathematically imposes a strict inequality between the mixed position-phason 2D Chern number in
Eq.~\eqref{eq:MixedC} and the 2D quantum metric in Eq.~\eqref{eq:OxOphi}, such that
\begin{equation}
    \Omega = \Omega_x+\Omega_\phi>\frac{|C|}{\pi}.
\end{equation}
Combining this inequality with the previously established identity between gap label and Chern number in 2D, $\nu = C$~\cite{kraus12,madsen13,jagannathan2025}, we obtain a {\it strict lower bound to the quantum metric} 
\begin{equation}
    \Omega>\frac{|\nu|}{\pi}\label{eq:minbound}.
\end{equation}
This strict bound on the trace of the quantum metric directly relates the localizability of the quasicrystal eigenstates, as measured by the quantum metric, to the energy gaps in the spectrum as quantified by the gap labels.
We note here that the quantum metric in real space is well-known to be the spread of the Wannier functions~\cite{marzari1997}, and as such directly encodes localization. In Sec.~\ref{sec:renorm} we further establish how the quantum metric also naturally captures the local symmetry centers in quasicrystals.

\begin{figure}
    \centering
    \includegraphics[width=.7\linewidth]{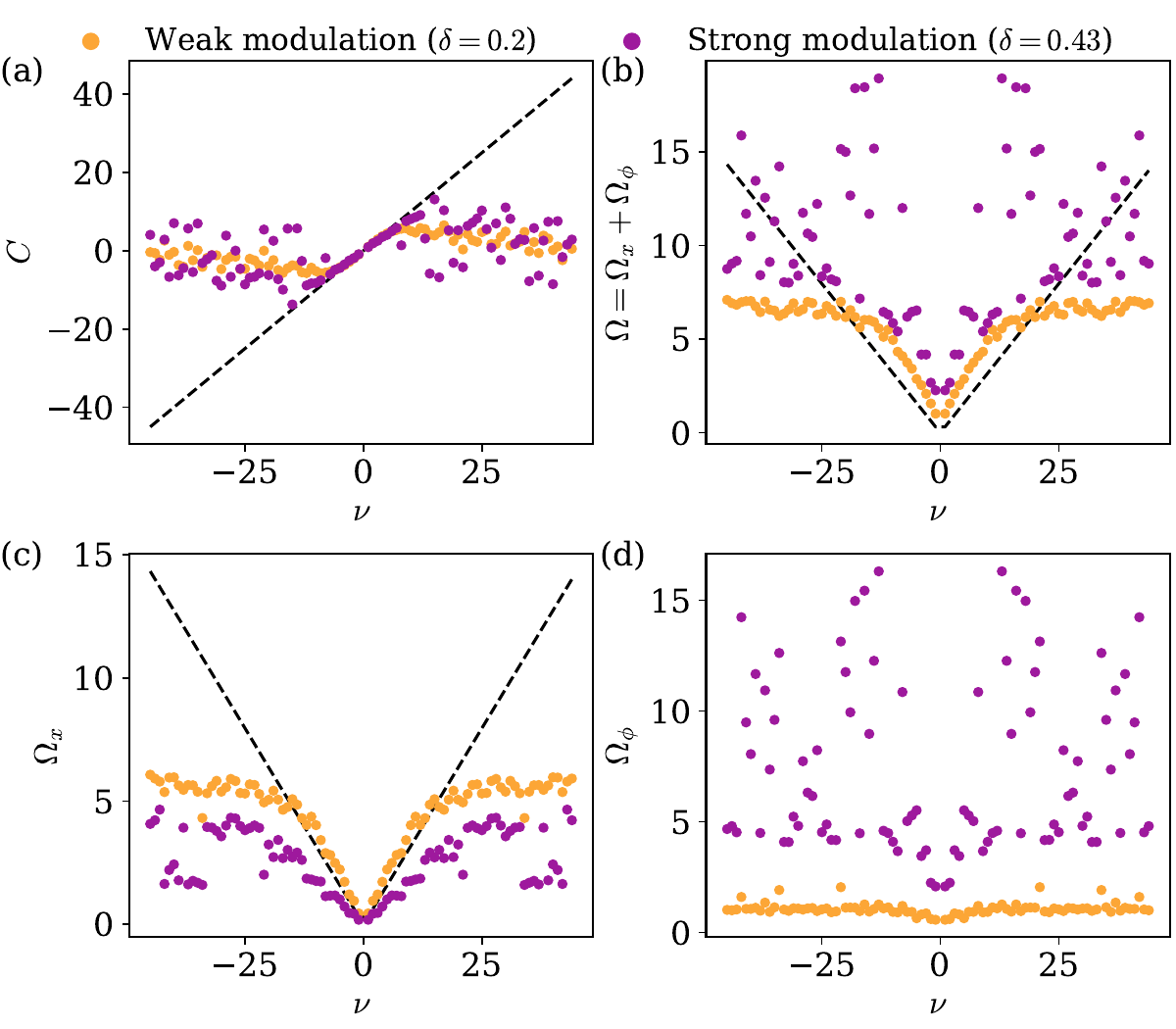}
    \caption{
    Chern number $C$ and quantum metric $\Omega = \Omega_x+\Omega_\phi$ as a function of the gap label $\nu$ with two colors corresponding to different modulation strengths $\delta = 0.2$, or $\delta = 0.43$, using  the 144-site approximant $F_{11}$ and OBCs, with a cutoff $L = N = 144$.
    (a) Chern number $C$, computed according to Eq.~\eqref{eq:MixedC}. Black dashed line indicates $C=\nu$.
    (b) Trace of the quantum metric $\Omega$, according to Eq.~\eqref{eq:OxOphi}. Black dashed line indicates $\Omega = \frac{|\nu|}{\pi}$.
    (c) Positional component $\Omega_x$ of the quantum metric, according to Eq.~\eqref{eq:Ox}. Black dashed line indicates $\Omega_x = \frac{|\nu|}{\pi}$.
    (d) Phasonic component $\Omega_\phi$ of the quantum metric, according to Eq.~\eqref{eq:Ophi}. }
    \label{fig:metrics}
\end{figure}

Having extracted strictly analytical lower bound on the quantum metric in Eq.~\eqref{eq:minbound}, we next turn to numerical investigations. In Fig.~\ref{fig:metrics}, we plot $C$, $\Omega = \Omega_x+\Omega_\phi$, as well as $\Omega_x$ and $\Omega_\phi$ separately, computed in the gaps of the approximant $F_{11}$, with a cutoff $L = N = 144$ for two different chain modulations $\delta$, as a function of the gap label $\nu$.
Small gap labels correspond to the largest energy gaps \cite{mace17} and for these gaps, the Chern number in Fig.~\ref{fig:metrics}(a) clearly converges to the gap label $C=\nu$ (dashed line).
This result confirms that the formulation introduced in Eq.~\eqref{eq:MixedC} is indeed able to characterize the topology in the mixed phason-position space. For larger gaps values, the Chern number as computed with Eq.~\eqref{eq:MixedC} deviates from the gap labels, as the edge states penetrate deeper in the bulk and bear opposite contributions Eq.~\eqref{eq:MixedC}~\cite{bianco2011}.

In Figs.~\ref{fig:metrics}(b-d) we plot both the total quantum metric $\Omega$ (b) and its positional $\Omega_x$ (c) and phasonic $\Omega_\phi$ (d) components and compare these to the Chern number. 
In Fig.~\ref{fig:metrics}(b) we immediately see that the trace of the quantum metric $\Omega$ scales linearly with $\nu$ both for weak and strong modulations, for all small gap labels, which have diminishing finite size effects. 
It is here interesting to note that we find numerically that a strict inequality between $\Omega$ and $\nu$ holds even for gaps where $C$ significantly deviates from $\nu$ due to finite size effects.

\begin{figure}
    \centering
    \includegraphics[width=0.9\linewidth]{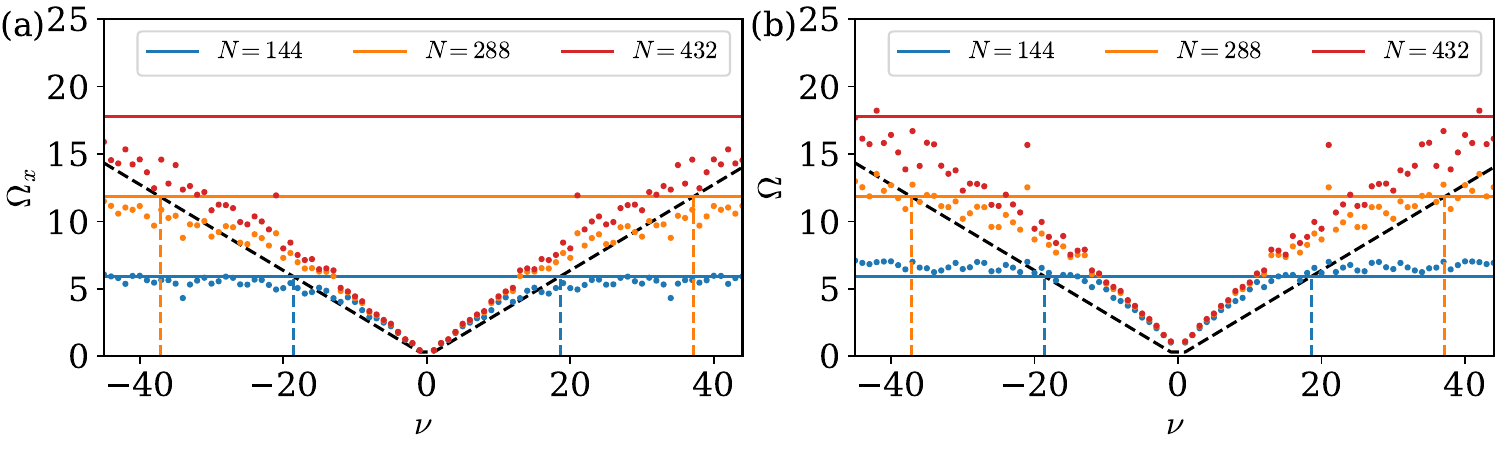}
    \caption{(a) Position component of the quantum metric $\Omega_x$ for $\delta = 0.2$ and different chain lengths $N$ (colors). Different chain lengths are obtained by concatenating one to three times the $F_{11}$ Fibonacci approximant, which contains $144$ sites. Horizontal lines show $4N/\pi^4$ for each length. Dashed vertical lines indicate $\nu_{max}$, for $N = 144$ and $N = 288$. For $N = 432$, $\nu_{max} \simeq 56$, so that no gap is affected by finite size effect. (b) Trace of the quantum metric $\Omega$ for the same approximants. 
    }
    \label{fig:gapmax}
\end{figure}

For large gap labels, however, the inequality Eq.~\eqref{eq:minbound} is clearly violated in the numerical results. 
We attribute the saturation of the quantum metric to the finite length of the chain: for short enough chains and large enough gap labels, equivalently small enough energy gaps, finite length effects will necessarily start to dominate over the effect of the quasicrystalline structure of the chain. 
In fact, we find that the finite length sets an {\it upper bound to the position quantum metric} in terms of the spatial spread of the eigenstates.
In App.~\ref{app:finite}, we analytically derive an estimate for this upper bound of the position part of the quantum metric as
\begin{eqnarray}
    \Omega_x<\Omega_0\approx\frac{4N}{\pi^4}.\label{eq:maxbound}
\end{eqnarray}
Here $\Omega_0$ is the quantum metric of the uniform chain, i.e.~for $\delta =0$, and $N$ is the length of the finite chain. We show that this upper bound is consistent with numerical results for the position quantum metric $\Omega_x$ in Fig.~\ref{fig:gapmax}(a), as we find $\Omega_x$ always bounded by $\nu_{max} = \frac{4N}{\pi^4}$ (horizontal colored lines). 
Moreover, we can utilize the bound in Eq.~\eqref{eq:minbound}, set by the positivity of the quantum geometric tensor, to extract a maximal gap label 
\begin{equation}
    \nu_{max} = \frac{4N}{\pi^3},\label{eq:numax}
\end{equation} 
beyond which finite size effects prevent $\Omega_x$ from scaling linearly with $\nu$. This maximal gap label is indicated by vertical dashed lines in Fig.~\ref{fig:gapmax}, showing full consistency between analytical derivation and numerical results.
However, the upper bound in Eq.~\eqref{eq:maxbound} cannot be directly applied to the trace of the quantum metric $\Omega$ as the phasonic component then also have to be taken into account. Still, our numerical results in Fig.~\ref{fig:gapmax}(b) show that $\nu_{max}$ remains relevant for the trace of the quantum metric $\Omega$, in the sense that for $|\nu|>\nu_{max}$, $\Omega$ saturates and Eq.~\eqref{eq:minbound} is then violated.

Finally, we discuss the dependence of the quantum metric on the modulation strength $\delta$. This dependence can be understood from the relative weight of its two components $\Omega_x$ and $\Omega_\phi$.
For non-zero but weak modulations $\delta\ll 1$, the chain becomes almost uniform and the phason dependence vanishes. Thus, the states delocalize along the chain, resulting in $\Omega_x \propto N$ the chain length, while $\Omega_\phi\xrightarrow[\delta\to0]{}0$. In this regime of small modulations, the trace of the quantum metric $\Omega$ thus coincides with the positional quantum metric $\Omega_x$ of the 1D Fibonacci chain.
As a consequence, for small enough modulations, numerically found to be typically $\delta<0.2$, the minimal localization of bulk states in the Fibonacci chain can be directly related to the gap label, according to
\begin{eqnarray}
    \Omega_x>\frac{|\nu|}{\pi}.
\end{eqnarray}
This result establishes that the localization properties, as measured by the positional quantum metric $\Omega_x$, must at least be linearly growing with the gap label $\nu$, which characterizes the energy spectrum.
Moreover, since the positional quantum metric $\Omega_x$ is not just bounded from below by the gap label, but also closely tracks it in this low modulation regime as seen in Fig.~ \ref{fig:metrics}(c), we conclude that the eigenstates of the Fibonacci chain inherit their localization properties from those of a 2D Chern insulator. This establishes a direct link between the spatial localization, characterized by the quantum metric, and the energy spectrum, as quantified by the gap label, in weakly modulated quasicrystals.
For strong modulations, $\delta\sim 1$, the phasonic part becomes more prominent in the trace of the quantum metric. The position component $\Omega_x$ still increases with gap label, but the strict inequality with the gap label does not hold anymore for only $\Omega_x$. We further note that we cannot deduce useful information for the 1D Fibonacci chain from the phasonic part alone as it is strongly influenced by the cutoff $L$. 
In fact, in the strong-modulation regime, the eigenstates delocalize along the phasonic dimension, resulting in a divergence in $\Omega_\phi$.

\subsection{Renormalization scheme and quantum metric}\label{sec:renorm}
In order to provide further understanding for both the strong relation between the quantum metric and the gap label and why the quantum metric is a such natural measure of localization in quasicrystals,
it is useful to discuss the renormalization group scheme that relates different approximants of the Fibonacci chain and their gap structure.
Due to the self-similarity properties of the Fibonacci chain, it contains many local symmetry centers, each having their own range of symmetry, {\it i.e.} the number of sites over which the symmetry is preserved.
In fact, each band of every approximant can be labeled by a set of mirror symmetry eigenvalues~\cite{piechon95, morfonios2014,mace17} representing how the states are transformed under every local mirror symmetry of the chain (see App.~\ref{app:fibo}).
The gap labels in turn keep track of this hierarchy of symmetry, since the smallest absolute labels separate bands with different symmetries at local scales, whereas larger gap labels, both positive and negative, only appear in long approximants, where long-range symmetries are present.

To proceed we first note that, the fact that the Fibonacci bands can be classified based on the local symmetries of the eigenstates they contain, makes the quantum metric a very natural probe of their localization.
Indeed, the quantum metric is maximally sensitive to having symmetric and antisymmetric states sharing the same symmetry center being either occupied or empty, and it is then proportional to the square of the spatial extent of such dimerized states. 
To explicitly illustrate how this is manifested in the Fibonacci chain, let us consider two groups of sites related by a spatial mirror symmetry. 
Conumbers sort sites according to their local environment (see App.~\ref{app:fibo}), so those sites have opposite conumbers: if one of the groups contains sites labeled $\{n_i\} = \{n_1, n_2, ...,  n_i\}$ in that order, then the sites of the symmetric group are labelled $\{-n_i\} = \{-n_i, ...,-n_2, -n_1\}$.
The spectrum of the Fibonacci chain then contains two states of opposite energy that are made of symmetric and antisymmetric superpositions of those groups of sites, which we may denote $\left|\pm\right> = (\left|\{n_i\}\right>\pm\left|\{-n_i\}\right>)/\sqrt{2}$.
Following Eq.~\eqref{eq:Ox}, this pair of states contributes to the positional quantum metric at all fillings where only one of them is occupied.
In fact, assuming, with no loss of generality, that $P^{(\phi)}\left|-\right> = \left|-\right>$ and $Q^{(\phi)}\left|+\right> = \left|+\right>$, then the following term appears in the sum of the quantum metric in Eq.~\eqref{eq:Ox}
\begin{multline}
    \left<-\right|x\left|+\right>\left<+\right|x\left|-\right> = \left|\left<+\right|x\left|-\right>\right|^2\\
    = \left(\left<\{n_i\}\right|x\left|\{n_i\}\right>-\left<\{-n_i\}\right|x\left|\{-n_i\}\right>\right)^2 = \left(\left<x\right>_{\{n_i\}}-\left<x\right>_{\{-n_i\}}\right)^2.\label{eq:superposition}
\end{multline}
Thus the positional quantum metric measures the spread, or extent, of the dimerized state between these two groups of sites related by a mirror symmetry.

Further, as shown in the App.~\ref{app:renorm}, each gap in the energy spectrum opens due to the hybridization of two similar groups of sites into symmetric and antisymmetric superpositions. Then Eq.~\eqref{eq:superposition} gives that the quantum metric calculated in that gap includes the measure of the square distance between these two groups of sites.
Thus, the smooth increase of the quantum metric as a function of increasing gap labels shown in Fig.~\ref{fig:metrics}(c) is directly related to the larger spatial separation in Eq.~\eqref{eq:superposition}.
Indeed, the gaps appearing deeper in the spectrum, with larger absolute gap labels, require longer approximants to appear and are associated to larger groups of sites. 
This establishes the quantum metric as a natural measure of localization and spatial structure of the eigenstates of quasicrystals~\footnote{An explicit derivation of the evolution of the quantum metric under renormalization group transformations has also been obtained independently after our work~\cite{wang2025}.}. This result is also consistent with previous results based on Kohn's localization length~\cite{varma2016}. 

In addition, our results also demonstrate that, by changing the filling fraction, such that the Fermi level changes between different gaps, it is possible to tune the localization length of the eigenstates to range from complete localization to a delocalization over the full length of the chain. Further tunability is also offered by changing the modulation strength of the chain.

\section{Conclusion}

In this work, we use the quantum metric to study the localization properties of the eigenstates in a quasicrystal, focusing on the 1D Fibonacci chain. 
The quantum metric probes the real-space spread of the electronic states.
In contrast to the IPR, often used for quasicrystals, it can thus efficiently discriminate states both occupying the same number of sites and with different separation in space, while still being an easily accessible observable. As we establish, the quantum metric even captures the importance of the many local symmetry centers present within the quasicrystal. Since such local symmetry centers are concatenated throughout the chain in a self-similar fashion, the quantum metric is in many ways an optimal and natural probe of locality on quasicrystals. With the quantum metric being numerically easily accessible and with experimental probes of it being rapidly uncovered \cite{verma2026}, the quantum metric thus offers great promise for understanding the physics of quasicrystals~\cite{carrasco2025}, both theoretically and experimentally.

Our results also demonstrate that the quantum metric is strongly connected to the many energy gaps present in quasicrystals. These gaps occur due to the hybridization between local symmetry-centered states, and thus the quantum metric is a natural quantity to connect structure with energy. In fact, building on the identity between the gap label and the Chern number~\cite{kraus12}, we establish a strict analytical lower bound on the quantum metric, consisting of the gap labels. The quantum metric is the sum of the positional quantum metric and an additional 'synthetic' dimension made of the phason paramter in quasicystals. We even find numerically that the quantum metric is closely tracking the gap label associated with each gap. Numerical results further establish that this bound remains relevant for the position component of the quantum metric alone for a range of modulation strengths, thus providing a direct link between spatial localization in a quasicrystal and its energy spectrum. 
We also demonstrate that this lower bound is violated when finite size effects dominate over fractality. We even establish analytically an estimate of an upper bound of the position component of the quantum metric in a finite chain length, along with an estimate of the maximal gap label below which it becomes relevant.
Taken together, our results establish that the quantum geometric tensor, composed of the Berry curvature and the quantum metric \cite{provost1980}, provides a comprehensive tool to understand many distinctive aspects of the quasicrystals.

Our results are connected to the fact that the localization properties of the bulk eigenstates in the 1D Fibonacci chain can be inferred from that of a 2D crystalline system.
The relation between properties of a 1D quasicrystal and a 2D crystal fundamentally stems from the fact that the 1D Fibonacci chain can be built by projecting a 2D square lattice on a line, also known as the cut-and-project method, see App.~\ref{app:fibo}.
This method is in fact a general method to obtain quasicrystals, in any dimension~\cite{katz1986, duneau1985,elser1986}. 
Thus, the approach we develop here to understand localization in the Fibonacci quasicrystal can be adapted to other quasicrystals, also in higher dimensions, whose spatial localization properties, expressed through the quantum metric, are then also inherited from higher-dimensional crystals, independent of them being topological or not.
This is reminiscent of the already establish relationship between the properties of quasicrystals and their higher-dimensional parent crystals in terms of their topological properties~\cite{kraus2016, else2021}, while our results establish it for the localization properties. Notably the quantum metric is still a useful concept even in non-topological systems.

Beyond the quest for understanding and characterizing the peculiar localization properties of quasicrystals, our work also emphasizes the quantum metric as an indispensable tool to quantify localization in self-similar systems and relate it to physical observables. 
Indeed, the quantum metric incorporates the notion of distance in real space, which is absent in many other tools currently used to study localization, such as the IPR. 
Moreover, the quantum metric can be directly related to easily accessible observables, such as conductivity~\cite{Wang2023} and optical responses~\cite{ozawa2019, li2021}, making it an experimentally accessible probe.
In the presence of electronic correlations, it has also been shown to be related to the fractional quantum Hall effect in Chern insulators \cite{bergholtz2013, ledwith2020, varjas2022} and the superfluid weight in flat-bands superconductors \cite{peotta2015, julku2016, tian2023}. This is an especially intriguing future direction, given the growing interest in superconducting quasicrystals~\cite{sakai2017,kamiya2018,zhang2022,wang2024,tokumoto2024,rai2019,rai2020,sandberg2024,fulga2016,ghadimi2021,kobialka2024,hori2024}.

\section*{Acknowledgements}
We thank A.~Bhattacharya, and A.~Jagannathan for interesting and fruitful discussions related to this work.

\paragraph{Author contributions}
Q. M. ran the calculations and wrote the initial version of the manuscript with inputs and help from P. H. A. B.-S. supervised and funded the project. All authors contributed to writing and revising the manuscript.

\paragraph{Funding information}
We acknowledge funding from the Swedish Research Council (Vetenskapr\aa det Grant No.~2022-03963) and the European Union through the European Research Council (ERC) under the European Union's Horizon 2020 research and innovation programme (ERC-2022-CoG, Grant Agreement No.~101087096). Views and opinions expressed are however those of the authors only and do not necessarily reflect those of the European Union or the European Research Council Executive Agency.

\begin{appendix}
\section{Methods}
In this Appendix we provide a summary of the methodology used in the main text for studying the Fibonacci chain and the quantum metric.
\subsection{Fibonacci chain}
\label{app:fibo}
\begin{figure*}
    \centering
    \includegraphics[width=1\linewidth]{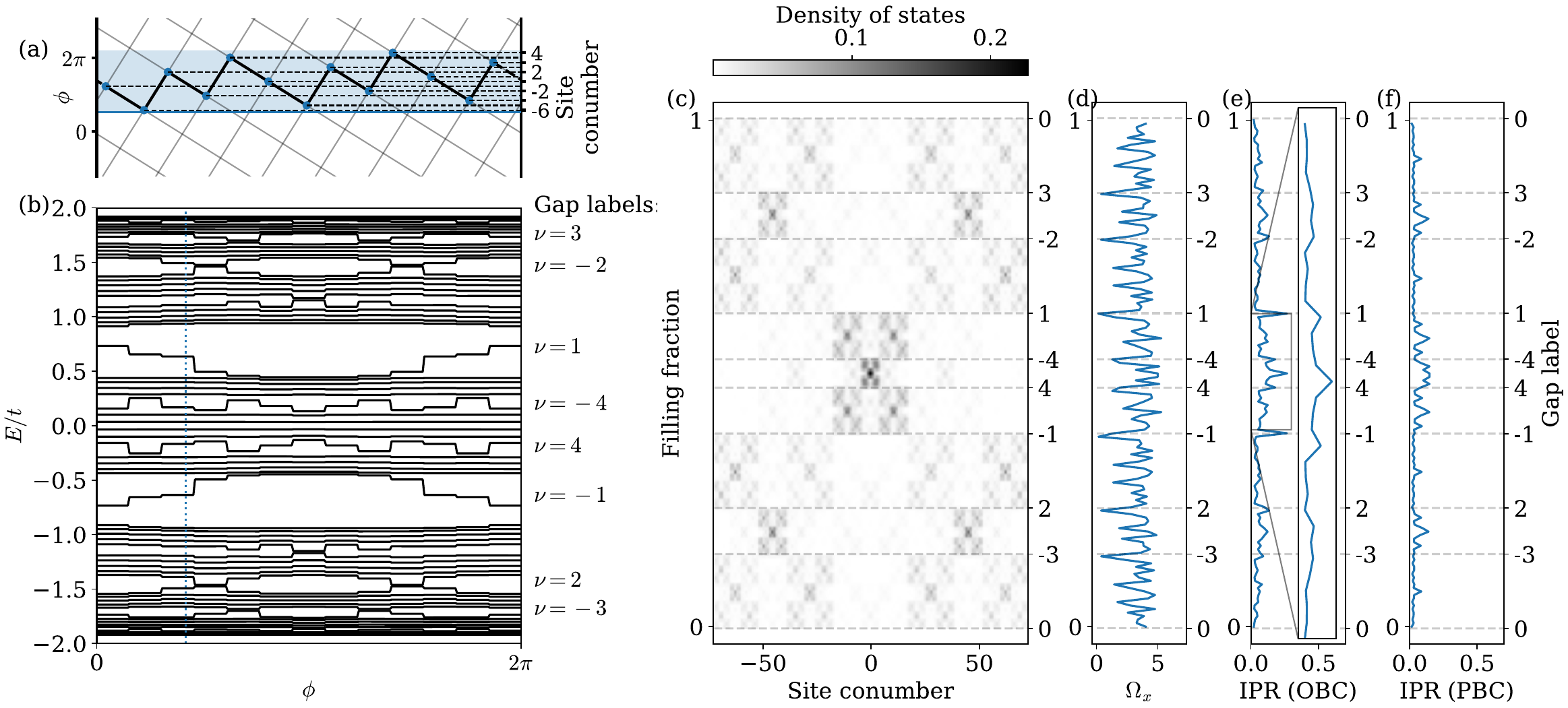}
    \caption{
    (a) Construction of a 13-site long periodic approximant of the Fibonacci chain by the `cut-and-project' method. The approximant is obtained by projecting a square lattice on the blue horizontal line along the $x$ direction. The slope of this line cut, here $5/8$, determines the length of the approximant.
    The vertical, $y$-axis, intercept of the line-cut determines the phason parameter $\phi$, see left $y$-axis. 
    Bonds along the $a$-axis project onto the horizontal line as long (L) bonds, while bonds along the $b$-axis project as short (S) bonds. 
    Sites can also be labeled by their conumber, an integer obtained by sorting sites by increasing projection on the orthogonal, $y$-axis, see right $y$-axis. 
    Sites with close conumbers have similar local environments. 
    (b) Energy spectrum obtained for the full range of phason parameter $\phi$ under OBCs. The spectrum acquires edge modes that wind through the energy gaps as a function of $\phi$. Gaps are labeled based on this winding number $\nu$. Dashed blue line marks spectrum for the phason mode drawn in (a).
    (c) Local density of states map of the eigenstates of the 144-site long Fibonacci approximant $F_{11}$ with modulation strength $\delta = 0.5$, averaged over the different phason modes. 
    Eigenstates are sorted by increasing filling fraction on the vertical $y$-axis, and sites are sorted by local environment, according to their conumber, along the $x$-axis. Dashed grey lines shows the position of the main energy gaps of the spectrum, along with their gap label on the right $y$-axis.
    Gap labels are also indicated on the right $y$-axis of panels (d), (e), (f).
    (d) Positional quantum metric $\Omega_x$ of the Fibonacci chain, as a function of filling fraction. The symmetry in the localization pattern between occupied and unoccupied bulk states results in a rich profile.
    (e) IPR of the eigenstates under OBCs, averaged over the phason modes. Inset shows a zoom-in in the middle of the spectrum to highlight the self-similar structure of the IPR. The presence of edge states results in sharp peaks in each gap.
    (f) IPR of the eigenstates under PBCs, averaged over the phason modes. In the absence of edge states, the IPR shows no peaks in the gap. The only variations reflect the number of sites on which each bulk state is localized.}
    \label{fig:chainfrac}
\end{figure*}

The Fibonacci chain, but also its finite approximants, can be built using the `cut-and-project' method~\cite{you88}, which highlights the long-range structure of the chain. 
It consists of cutting a stripe from a tilted 2D square lattice, and projecting the sites it contains onto a 1D line, as illustrated in Fig.~\ref{fig:chainfrac}(a). 
When projected along the line-cut direction, the bonds along the $a$-axis result in long bonds $t_L$, whereas bonds along the $b$-axis project as short bonds $t_S$.
The slope of the tilted square lattice, $\tan(\alpha)$, entirely determines the obtained chain: rational slopes correspond to periodic chains, while quadratically irrational slopes give rise to true quasicrystals. 
The Fibonacci chain corresponds to a slope equal to the inverse golden mean $\tan(\alpha) = \frac{2}{1+\sqrt{5}}$. 
Slopes given by ratios of consecutive terms of the Fibonacci sequence ($3/2$, $5/3$, $8/5$, $13/8$...) result in finite sections of the full Fibonacci chain repeating periodically. 
Such sequences are called finite approximants and labelled $F_n$, $n\in \mathbb{N}$.

The `cut-and-project' method has the benefit that projecting sites along the perpendicular direction to the line cut, {\it i.e.}~vertical axis in Fig.~\ref{fig:chainfrac}(a), allows us to associate a so-called conumber to each site, an integer labeling sites based on their vertical position~\cite{jagannathan21}. 
The conumber reflects the local environment of each site, with sites with close conumbers sharing similar local environments and sites with opposite conumbers having mutually symmetric local environments~\cite{sire89, rontgen19}.
In the example shown in Fig.~\ref{fig:chainfrac}(a), the approximant contains $13$ sites, with sites thus conumbered from $-6$ to $6$.
Sites with conumber between $-6$ and $-2$ have a long bond on their left and a short bond on their right, whereas the opposite is true for sites with conumbers between $2$ and $6$. 
All these sites form pairs around a strong $t_S$ bond and can therefore be called molecular sites.
Sites with conumbers $-1$, $0$, and $1$ instead sit in-between long bonds. These sites thus become isolated in the strongly modulated limit, $t_L\ll t_S$, and can therefore be called atomic sites~\cite{wang2024}.

The sequence of hopping terms obtained by the `cut-and-project' method also depends on the $y$-intercept of the line cut, which can be labeled by the so-called phason parameter, or angle, $\phi$.
One can therefore write the Hamiltonian as
\begin{equation}
    \mathcal{H}(\phi) = \sum_n -t_n(\phi)c_n^\dagger c_{n+1} + {\rm H.c.} \label{eq:hamphi}, 
\end{equation}
where the explicit dependence of $t_n$ on $\phi$ is given by Eq.~\eqref{eq:ph_dep}.
If we here promote $\phi$ to an independent degree of freedom for the system, we obtain Eq.~\eqref{eq:Hphi2}.
Here, Eq.~\eqref{eq:ph_dep} shows that a shift in $\phi$ is equivalent to a shift of the site position $n$ and can thus be discarded by a change of the origin of positions.
This makes the phason parameter irrelevant under periodic boundary conditions (PBCs).
However, under open boundary conditions (OBCs) different phason parameters boil down to opening the chain at different sites, and thus, with different terminations.
Depending on the termination, edge states may form in the energy gaps between bands of energy states of the PBC chain~\cite{wehling2010}. Note that bands here designate the ranges of energy in which the eigenstates lie for different phason parameters, but there is no regular dispersion relation as found in periodic systems.
We illustrate this in Fig.~\ref{fig:chainfrac}(b) by plotting the energy spectrum as a function of the phason parameter $\phi$. Clear gaps are identified in the spectrum. 
As $\phi$ sweeps over the different phason modes, the edge states wind within these gaps~\cite{rontgen19}. 
This winding of the edge states can be understood based on an analogy between one-dimensional quasicrystals and 2D Chern insulators~\cite{kraus12, kraus2012b, madsen13,jagannathan2025}, mapping $\phi$ to a momentum-like degree of freedom.
Thanks to this analogy, the energy gaps under PBCs can be labeled according to the winding number $\nu$ of the edge states inside each gap, as indicated in Fig.~\ref{fig:chainfrac}(b).
This so-called gap label can also be determined analytically based on the integrated density of states~\cite{mace17}.
It is worth noting that gaps with higher winding numbers $\nu$ are smaller. 
This implies that the edge states winding in those gaps penetrate deeper in the bulk compared to edge states in larger gaps, which have smaller (absolute valued) gap labels. 
Thus, longer approximants are necessary to correctly resolve the narrowest gaps.

\subsection{Quantum metric}
\label{app:metric}
The quantum geometric tensor is an observable of wave function geometry in the Hilbert space.
Its real part, the quantum metric, measures the volume of the Hilbert space spanned by a given band, or other set of states~\cite{provost1980}.
In crystalline systems, since the momentum $k$ is then a proper quantum number, the quantum metric is usually expressed using derivatives of the wave function
\begin{equation}
    \Omega = \frac{1}{2\pi}\int\mathrm{d} k \mathrm{Tr}\left[\partial_k P(k)(1-P(k))\partial_kP(k)\right]\label{eq:Ok},
\end{equation}
where $P(k)$ is the projector on the set of occupied bands at momentum $k$.
However, due to the absence of translational invariance in the Fibonacci chain, the eigenstates cannot be labeled by momentum.
We thus need an equivalent formulation in position space. 
Such a formulation has already been introduced for 2D systems~\cite{marzari1997, marsal2024}.
Here, we extend this formulation to 1D quasicrystals, to be able to apply it to the Fibonacci chain:
\begin{equation}
    \Omega = \frac{1}{2\pi N}\int_0^{2\pi}\mathrm{d}\phi\mathrm{Tr}\left[P^{(\phi)}xQ^{(\phi)}xP^{(\phi)}\right]\label{eq:metricpos},
\end{equation}
where $P^{(\phi)}$ is the projector on the occupied states of the phason mode $\phi$, $Q^{(\phi)} = 1-P^{(\phi)}$, and $x$ is the position operator, while $N$ is the number of sites in the chain.
To be able to define the position operator $x$, the chain needs to have OBCs, because under PBCs, the position operator cannot avoid an artificial sudden jump from $N$ to $0$ somewhere along the chain.
Since we are interested in the properties of bulk states, and thus want to avoid that they depend on the chain termination, we use the phason-averaged quantum metric in Eq.~\eqref{eq:metricpos}.

Equation~\eqref{eq:metricpos} gives the quantum metric a simple interpretation. 
It measures the second moment of the electron distribution in position space, {\it i.e.}~the spatial extension of the maximally localized Wannier orbitals~\cite{marzari1997}, for a given set of states. 
Indeed,
\begin{eqnarray}
    \mathrm{Tr}\left[PxQxP\right] &=& \mathrm{Tr}\left[Px^2P\right]-\mathrm{Tr}\left[PxPxP\right]\\
    &=& \left<x^2\right>-\left<x\right>^2,
\end{eqnarray}
where $\left<\cdot\right>$ is the expectation value over the set of occupied states.
Thus, the quantum metric directly encodes the distance between the sites on which the eigenstates are localized~\cite{marsal2024}. 
From Eq.~\eqref{eq:metricpos} we also see that the strongest contributions to the quantum metric come from pairs of occupied and unoccupied states lying on the same sites, ensuring that $P^{(\phi)}xQ^{(\phi)}\neq0$. 
Figure~\ref{fig:chainfrac}(c) shows that this is precisely the case for the Fibonacci chain. 
It shows the local density of each state (y-axis) as a function of the local environment, indicated by the site conumber on the horizontal x-axis.
It has a very symmetric structure, showing that, whatever the filling fraction we consider, we can always find occupied and unoccupied bulk states on the same groups of sites, resulting in the rich and self-similar profile for the quantum metric showed in Fig.~\ref{fig:chainfrac}(d).
On the contrary, the $\phi$-dependent edge states contribute very little to the quantum metric since edge states in gaps at opposite energy lie at opposite end of the chain, and thus have no overlap.

\subsection{Inverse particiation ratio (IPR)}
\label{app:IPR}
The quantum metric in Fig.~\ref{fig:chainfrac}(d) is in sharp contrast with traditional measures of localization, such as the inverse participation ratio (IPR).
The IPR is defined for each eigenstate $\left|\psi\right>$, as $\mathrm{IPR}(\left|\psi\right>) = \sum_{\mathrm{sites}\ i}\left|\left<i|\psi\right>\right|^4$.
It can intuitively be interpreted as the expectation value of the inverse number of sites on which a given state is localized. It therefore ranges from $0$ for a  state with equal amplitude on every site of an infinite system, {\it i.e.}~fully delocalized, to $1$ for a state fully localized on a single site. It has therefore been proven to be a useful tool to study the localization properties of eigenstates~\cite{moustaj2021,ahmed2022}. 
However, the IPR only takes into consideration the number of occupied sites, thereby not retaining any information about the structure of the chain. 
As such, it is also mainly sensitive to the localization of the edge states, showing sharp peaks under OBCs as seen in Fig.~\ref{fig:chainfrac}(e), which disappear under PBCs in Fig.~\ref{fig:chainfrac}(f). Specifically, the IPR seems unable to discriminate the localization properties of the different bulk states, which is crucial for understanding the structure of the Fibonacci chain.

\subsection{Position space renormalization scheme}
\label{app:renorm}
We here illustrate the symmetries of the Fibonacci chain using a real-space renormalization-group approach. 
For that, it is useful to note that the quasicrystalline sequence of hopping terms $(t_n)$ can be obtained from successive concatenations of longer and longer approximants, following the Fibonacci recursion relation. This is a fully equivalent construction fo the Fibonacci chain to the `cut-and-project' method~\cite{jagannathan21}.
The two shortest, trivial, approximants are $F_0 = t_S$ and $F_1 = t_L$, and then with all others obtained through the concatenations $F_n = F_{n-2}F_{n-1}$. 
Thus, the shortest non-trivial approximants are
\begin{eqnarray}
    F_2 &=& t_St_L,\\
    F_3 &=& t_Lt_St_L, \\
    F_4 &=& t_St_Lt_Lt_St_L,\\
    F_5 &=& t_Lt_St_Lt_St_Lt_Lt_St_L, \\
    &\textit{etc.}&
\end{eqnarray}
This recursive construction makes the Fibonacci chain automatically self-similar. 
For example, $F_2$ appears three times in $F_5$, $F_3$ appears twice, and $F_4$ once. 
The self-similarity is also seen from the fact that each approximant can be obtained from the previous, shorter one using the inflation rules~\cite{jagannathan21}
\begin{eqnarray}
    t_S &\rightarrow& t_L,\\
    t_L &\rightarrow& t_St_L.
\end{eqnarray}
As a direct consequence, the Hamiltonian, and thus the energy spectrum and wave functions, also become self-similar, as illustrated in Fig.~\ref{fig:chainfrac}.

\begin{figure}[tb]
    \centering
    \includegraphics[width=.6\linewidth]{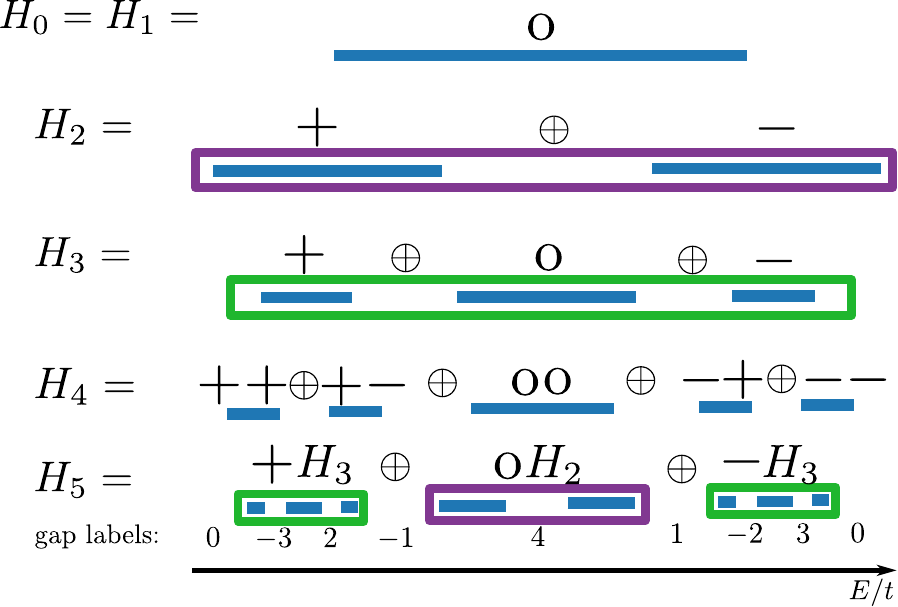}
    \caption{Schematic band structure of the six smallest Fibonacci approximants along with their symmetry eigenvalues. 
    Blue lines are illustrations of the band structure of each approximant, but with length not to scale. 
    To each band corresponds a set of symmetry eigenvalues representing atomic (o), molecular symmetric ($+$) and molecular antisymmetric ($-$) characters, with $H_n$ being the set of symmetry eigenvalues of the bands obtained from the approximant $F_n$. The symmetry eigenvalues of different (larger) bands are separated by $\oplus$. 
    The two trivial approximants $F_0$ and $F_1$ share the same spectrum. Then, $F_2$ is the SSH chain, with two bands. The spectra of the following approximants are obtained from Eq.~\eqref{eq:renorm}. Gap labels of $F_5$ are also shown.}
    \label{fig:renorm}
\end{figure}

Figure~\ref{fig:renorm} summarizes the band structures of the first few approximants of the Fibonacci chain under PBCs, with the corresponding local symmetry eigenvalues, with $H_n$ being the set of symmetry eigenvalues of the bands obtained for approximant $F_n$.
The shortest approximants $F_0$ and $F_1$ have just one type of bonds, and thus form only one band.
All sites are thus local symmetry centers of the chain and the band can be said to have an atomic symmetry, which we denote by "o".
The next approximant $F_2$ corresponds to the SSH polyacetylen chain~\cite{ssh1979}.
In this approximant, the local symmetry centers lie on the bond, and the spectrum thus decomposes into two molecular bands lying on a dimer, one symmetric ($+$) and one antisymmetric ($-$) with respect to the bond center. 
The first non-trivial approximant $F_3$ (under PBC) hosts two kinds of local environments: atomic sites ($t_L\cdot t_L$ chain segments), and molecular dimers ($t_L\cdot t_S\cdot t_L$).
Thus, its spectrum decomposes into an atomic band (o) localized on atomic sites, flanked in energy by the molecular symmetric ($+$) and antisymmetric ($-$) bands occupying dimers.

It is now interesting to note that longer approximants retain the main three-band structure found in the first four approximants. 
This directly stems from the self-similarity of the Fibonacci chain and indicates that the eigenstates of all approximants can be characterized by a local symmetry eigenvalue: $+$, $-$ or o.
As we construct longer approximants, the three main bands also split into sub-bands, that can be associated to symmetries emerging at longer scales. 
For example, $F_5$ hosts the same three types of local environments as $F_3$, but we can further distinguish sites based on their larger environment: dimers can be either grouped as pairs when they are separated by a single long bond ($t_L t_L\cdot t_S\cdot t_L\cdot t_S\cdot t_Lt_L$) or isolated if they are only surrounded by long bonds ($t_Lt_L\cdot t_S\cdot t_Lt_L$); while atomic sites all get grouped as pairs separated by a dimer ($t_L\cdot t_L t_S t_L \cdot t_L$).
As a consequence, the atomic band (o) inherits long-range symmetries from $F_2$ (purple rectangle in Fig.~\ref{fig:renorm}) and thus splits into atomic symmetric (o$+$) and atomic antisymmetric (o$-$) bands. 
Similarly, the molecular bands ($\pm$) split into three sub-bands, inheriting long-range symmetries from $F_3$ (green rectangles in Fig.~\ref{fig:renorm}). 
They form two molecular-atomic ($+$o, $-$o) and four molecular-molecular ($++$, $+-$, $-+$, $--$) bands. With the energy splitting being lower for the atomic bands, this together results in the structure shown for $F_5$ in Fig.~\ref{fig:renorm}.

When generalizing the scheme displayed in Fig.~\ref{fig:renorm} to ever larger approximants, their sub-band structure can be obtained from shorter approximants after performing a step of the renormalization scheme, seen in Fig.~\ref{fig:renorm}. 
Indeed, $H_n$ for any $n$ can be obtained from the recursion relation~\cite{mace17, mace2016},
\begin{eqnarray}
    H_0 &=& H_1 = \mathrm{o},\\
    H_2 &=& +\oplus -,\\
    H_n &=& +H_{n-2}\oplus \mathrm{o}H_{n-3}\oplus -H_{n-2}\label{eq:renorm},
\end{eqnarray}
where the symmetry labels of each band (o, $\pm$) are ordered by smaller local atomic environment on the left to larger ones on the right. 
This hierarchy of symmetries at larger and larger scale is consistent with the picture of requiring larger and larger local neighborhoods to resolve states around ever narrower gaps, also resulting in edge states penetrating far into the bulk.

\section{Fourier transform of the phasonic degree of freedom}\label{app:FT}
In this Appendix, we provide details for the derivation of Eq.~(8) and (9) in the main text. This calculation was already published earlier in~Ref.~\cite{kraus12}, although with some typos.
We first recall that the hopping function
\begin{equation}
    t_n(\phi) = \frac{t_L+t_S}{2}(1+\delta\ \mathrm{sgn}[\cos\left(2n\pi \tan(\alpha)+\phi\right)-\cos\left(\pi\tan(\alpha)\right)]),
\end{equation}
is a periodic function of $\phi$. 
Thus, it can be decomposed into a Fourier series as in Eq. (7) of the main text, where for $n\in \mathbb{Z}$, the $n$-th Fourier coefficient $\tilde{t}_n$ reads
\begin{eqnarray}
    \tilde{t}_n^l &=& \frac{1}{2\pi}\int_{-\pi}^{\pi}t_n(\phi)e^{il\phi}\mathrm{d}\phi,\\
    &=&\frac{1}{2\pi}\frac{t_L+t_S}{2}\left(2\pi\delta_{l0}+\delta\int_{-\pi}^{\pi}\mathrm{sgn}[\cos(\frac{2\pi nq}{p+q}+\phi)-\cos(\frac{\pi q}{p+q})]e^{il\phi}\mathrm{d}\phi \right)\\
    &=&\frac{t_L+t_S}{2}\left(\delta_{l0}+\frac{\delta e^{-2i\pi nl\frac{q}{p+q}}}{2\pi}\int_{-\pi}^{\pi}\mathrm{sgn}[\cos(\phi)-\cos(\frac{\pi q}{p+q}))]e^{il\phi}\mathrm{d}\phi\right)\\
    &=&\frac{t_L+t_S}{2}\left(\delta_{l0}+\frac{\delta e^{-\frac{2i\pi nlq}{p+q}}}{2\pi}\left[\int_{-\pi}^{-\frac{\pi q}{p+q}}-e^{il\phi}\mathrm{d}\phi + \int_{-\frac{\pi q}{p+q}}^{\frac{\pi q}{p+q}}e^{il\phi}\mathrm{d}\phi + \int_{\frac{\pi q}{p+q}}^{\pi}-e^{il\phi}\mathrm{d}\phi\right]\right).
\end{eqnarray}
We then note that $\frac{q}{p+q}$ is a rational approximant of the inverse golden ratio $\frac{2}{1+\sqrt{5}}$, so that $\frac{\pi}{2}\leq\pi\frac{q}{p+q}<\pi$ for any approximant of the Fibonacci chain.
Hence, for $l = 0$ we find that
\begin{equation}
    \tilde{t}_n^0 = \frac{t_L+t_S}{2}\left(1+\delta\left(2\frac{q}{p+q}-1\right)\right),
\end{equation}
while for $l\neq0$ we find that
\begin{eqnarray}
    \tilde{t}_n^l &=& \delta\frac{t_L+t_S}{2}e^{-\frac{2i\pi nlq}{p+q}}\frac{1}{2i\pi l}\left[e^{-il\pi}-e^{-il\frac{\pi q}{p+q}}+e^{\frac{il\pi q}{p+q}}-e^{\frac{-il\pi q}{p+q}}-e^{il\pi}+e^{\frac{il\pi q}{p+q}}\right]\\
    &=&\frac{t_S-t_L}{2}e^{\frac{-2i\pi nlq}{p+q}}\frac{2\sin(\frac{\pi ql}{p+q})}{\pi l}.
\end{eqnarray}
Here, $\tilde{t}_n^l$ can be interpreted as a hopping to the $l$-th neighbor along the conjugate variable to the phasonic degree of freedom. 
Note that the sharp phason flip as a function of $\phi$ results in long-range hoppings $l$ in this conjugate space.

\section{Finite size effects}\label{app:finite}
In this Appendix, we discuss the effect of the finite size of the chain on the calculations of the quantum metric in Fig.~\ref{fig:metrics} and Fig.~\ref{fig:gapmax} and derive Eq.~\eqref{eq:maxbound} and Eq.~\eqref{eq:numax} of the main text.
In particular, we compute the quantum metric $\Omega_0$ of a finite uniform chain, {\it i.e.} with $\delta = 0$. 
We find that it remains nearly independent of the filling fraction, and only depends on the length of the chain. 
Based on the numerical observation that a finite modulation strength always decreases the position component of the quantum metric in the Fibonacci chain, compared to the uniform case $\delta = 0$, we can deduce an upper bound to the $\Omega_x$ at finite $\delta$.

We use this bound to analytically define the sets of gap labels for which Eq.~\eqref{eq:minbound} holds true. 

\begin{figure}
    \centering
    \includegraphics[width=.4\linewidth]{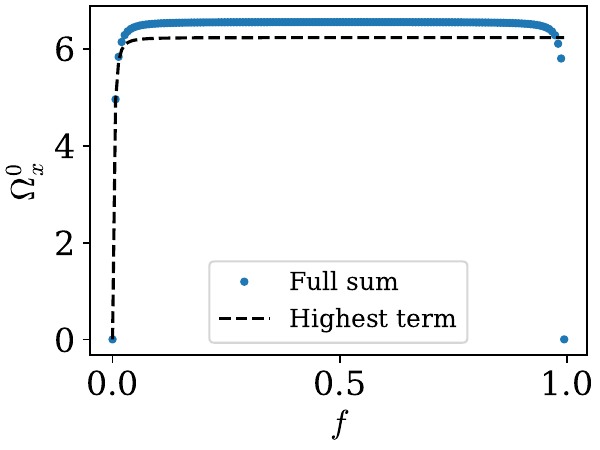}
    \caption{(a) Quantum metric $\Omega_{0}$ of a finite uniform chain with $N = 150$ sites as a function of the filling fraction $f$. Dots represent the numerical evaluation of the sum in Eq.~\eqref{eqapp:sum}, while dashed line represents its approximation by the highest term.
    }
    \label{fig:finite}
\end{figure}

In the case of vanishing modulation strength $\delta=0$, the chain just has a uniform hopping strength $t_0$. 
The different phason realizations thus become identical and $\Omega_\phi$ vanishes. 
Thus, $\Omega_0 = \Omega_{0,x}$, i.e.~the positional component of the quantum metric.
Under open boundary conditions, the Hamiltonian of a uniform chain made of N sites reads $H = \sum_{n=1}^{N-1}t_0c_n^\dagger c_{n+1}$.
The eigenstates are conveniently expressed as
\begin{eqnarray}
    \psi_k(n) = \sqrt{\frac{2}{N+1}}\sin\left(\frac{k\pi n}{N+1}\right),\quad k\in[1,N],
\end{eqnarray}
with corresponding energy
\begin{eqnarray}
    E_k = 2t_0\cos\left(\frac{k\pi n}{N+1}\right).
\end{eqnarray}
Knowing both the eigenstates and eigenvalues, we can compute $\Omega_0 = \Omega_{0,x}$ at any filling fraction $f = \frac{k_0}{N}$.
\begin{eqnarray}
    \Omega_{0} &=& \frac{1}{2\pi N}\int_0^{2\pi} \mathrm{Tr}[P^{(\phi)}xQ^{(\phi)}xP^{(\phi)}]\mathrm{d}\phi\\
    &=&\frac{1}{N}\mathrm{Tr}[P^{(0)}xQ^{(0)}xP^{(0)}]\\
    &=& \frac{1}{N}\sum_{k =1}^{k_0}\sum_{k' = k_0+1}^{N}\left(\sum_{n = 0}^{N}n\psi_k(n)\psi_{k'}(n)\right)^2,
\end{eqnarray}
where we use that $P^{(\phi)}$ and $Q^{(\phi)}$ are independent of $\phi$ when $\delta = 0$. Note that, even if sites are labeled from $1$ to $N$ in the definition of the Hamiltonian, we may extend the sum over sites $n$ to the range $[0,N]$ in order to simplify notations, since $\psi_k(0) = 0$ for any $k$.
The sum over $n$ in the parenthesis can be evaluated using Riemann series:
\begin{eqnarray}
    \sum_{n = 0}^{N}n\psi_k(n)\psi_{k'}(n) &=& \frac{2}{N+1}\sum_{n=0}^N n \sin\left(\frac{k\pi n}{N+1}\right)\sin\left(\frac{k'\pi n }{N+1}\right)\\
    &=& \frac{1}{N+1}\sum_{n=0}^N n \left(\cos\frac{(k-k')\pi n}{N+1}-\cos\frac{(k+k')\pi n}{N+1}\right)\\
    &\approx& \frac{1}{N+1}(N+1)^2\int_0^1x\left(\cos(k-k')\pi x-\cos(k+k')\pi x\right)\\
    &\approx& \frac{N+1}{\pi^2}\left(\frac{2}{(k-k')^2}-\frac{2}{(k+k')^2}\right)\\
    &\approx& \frac{N+1}{\pi^2}\frac{8kk'}{(k^2-k'^2)^2}.
\end{eqnarray}
Hence the quantum metric reads
\begin{eqnarray}
    \Omega_{0} \approx \frac{64(N+1)^2}{N\pi^4}\sum_{k=0}^{k_0}\sum_{k'=k_0+1}^N\left(\frac{kk'}{(k^2-k'^2)^2}\right)^2.\label{eqapp:sum}
\end{eqnarray}
This sum is difficult to compute analytically.
However, by a change of variable $(k,k')\mapsto(d=k-k', s = (k+k')/2)$, we can rewrite this sum as partial sum of a series whose general terms decay as $O(1/d^3)$. 
In order to obtain a simple numerical estimate, we approximate this sum by keeping only the term $d=1$, obtained for $k=k_0$ and $k' = k_0+1$
\begin{equation}
    \Omega_{0} \approx \frac{64(N+1)^2}{N\pi^4}\frac{k_0^2(k_0+1)^2}{(k_0^2-(k_0+1)^2)^4} = \frac{64(N+1)^2}{N\pi^4}\frac{k_0^2(k_0+1)^2}{(2k_0+1)^4}.\label{eqapp:Omega_unif}
\end{equation}
 This approximation boils down to keeping only the contribution from the pairs of occupied and unoccupied states that are closest in energy.
Figure~\ref{fig:finite}(a) shows the discrepancy introduced by this approximation as a function of the filling fraction $f=k_0/N$ for a chain with $N = 150$ sites.
From Eq.~\eqref{eqapp:Omega_unif}, we obtain the scaling of $\Omega_0$ in $N$, as keeping $k_0/N = f$ constant,
\begin{equation}
    \Omega_{0}\propto \frac{64 N}{\pi^4}\times\frac{f^2f^2}{(2f)^4} = \frac{4N}{\pi^4}.
\end{equation}
Interestingly, the position quantum metric of a 1D uniform chain is thus mostly independent of the filling fraction.
This can be confirmed by a numerical evaluation of the sum in Eq.~\eqref{eqapp:sum}, as illustrated in Fig.~\ref{fig:finite}(a).
The approximation only breaks down for filling fraction close to 0 or 1.

Coming back to the Fibonacci chain, with $\delta>0$, the eigenstates can never be more delocalized than the Bloch wave of the uniform chain. Thus, $\Omega_0$ sets an upper bound to $\Omega_x$:
\begin{equation}
    \Omega_x<\Omega_0 \approx \frac{4N}{\pi^4}.
\end{equation}
This is confirmed by our numerical calculations, showed in Fig.~\ref{fig:gapmax}(a).
As a direct consequence, the linear scaling of $\Omega_x$ with $\nu$ predicted in Eq.~\eqref{eq:minbound} can only by maintained as long as
\begin{equation}
    \frac{|\nu|}{\pi}<\frac{4N}{\pi^4},\label{eqapp:finitesize}
\end{equation}
{\it i.e.} for $|\nu|<\nu_{max}$ with 
\begin{equation}
    \nu_{max} = \frac{4N}{\pi}.
\end{equation}
In order to crosscheck this result against a numerical evaluation, we compare the quantum metric and its position component obtained for 1, 2 and 3 repetitions of the approximant $F_11$ in Fig.~\ref{fig:gapmax}.
This corresponds to chains made of $N = 144, 288$, and $432$ sites, respectively. 
According to Eq.~\eqref{eqapp:finitesize}, these chains can be used to probe gaps $|\nu|\lesssim \nu_{max}$, with $\nu_{max} = \frac{4N}{\pi^3} = 18, 37, 56$ respectively.
Note that these limits seem to be independent of the modulation strength, as also seen in Fig~\ref{fig:metrics}.
This means that the finite size effects are independent of the actual gap width in the energy spectrum.

\end{appendix}



\begin{thebibliography}{10}
\providecommand{\url}[1]{\texttt{#1}}
\providecommand{\urlprefix}{URL }
\expandafter\ifx\csname urlstyle\endcsname\relax
  \providecommand{\doi}[1]{doi:\discretionary{}{}{}#1}\else
  \providecommand{\doi}{doi:\discretionary{}{}{}\begingroup \urlstyle{rm}\Url}\fi
\providecommand{\eprint}[2][]{\url{#2}}

\bibitem{Shechtman1984}
D.~Shechtman, I.~Blech, D.~Gratias and J.~W. Cahn,
\newblock \emph{Metallic phase with long-range orientational order and no translational symmetry},
\newblock Phys. Rev. Lett. \textbf{53}, 1951 (1984),
\newblock \doi{10.1103/PhysRevLett.53.1951}.

\bibitem{mace2017b}
N.~Mac\'e, A.~Jagannathan, P.~Kalugin, R.~Mosseri and F.~Pi\'echon,
\newblock \emph{Critical eigenstates and their properties in one- and two-dimensional quasicrystals},
\newblock Phys. Rev. B \textbf{96}, 045138 (2017),
\newblock \doi{10.1103/PhysRevB.96.045138}.

\bibitem{kraus2012b}
Y.~E. Kraus, Y.~Lahini, Z.~Ringel, M.~Verbin and O.~Zilberberg,
\newblock \emph{Topological states and adiabatic pumping in quasicrystals},
\newblock Phys. Rev. Lett. \textbf{109}, 106402 (2012),
\newblock \doi{10.1103/PhysRevLett.109.106402}.

\bibitem{varjas2019}
D.~Varjas, A.~Lau, K.~P\"oyh\"onen, A.~R. Akhmerov, D.~I. Pikulin and I.~C. Fulga,
\newblock \emph{Topological phases without crystalline counterparts},
\newblock Phys. Rev. Lett. \textbf{123}, 196401 (2019),
\newblock \doi{10.1103/PhysRevLett.123.196401}.

\bibitem{rai2021}
G.~Rai, H.~Schl\"omer, C.~Matsumura, S.~Haas and A.~Jagannathan,
\newblock \emph{Bulk topological signatures of a quasicrystal},
\newblock Phys. Rev. B \textbf{104}, 184202 (2021),
\newblock \doi{10.1103/PhysRevB.104.184202}.

\bibitem{else2021}
D.~V. Else, S.-J. Huang, A.~Prem and A.~Gromov,
\newblock \emph{Quantum many-body topology of quasicrystals},
\newblock Phys. Rev. X \textbf{11}, 041051 (2021),
\newblock \doi{10.1103/PhysRevX.11.041051}.

\bibitem{fan-huang2022}
J.~Fan and H.~Huang,
\newblock \emph{Topological states in quasicrystals},
\newblock Front. Phys. \textbf{17}(1), 13203 (2022),
\newblock \doi{10.1007/s11467-021-1100-y}.

\bibitem{mace2019}
N.~Mac\'e, N.~Laflorencie and F.~Alet,
\newblock \emph{{Many-body localization in a quasiperiodic Fibonacci chain}},
\newblock SciPost Phys. \textbf{6}, 050 (2019),
\newblock \doi{10.21468/SciPostPhys.6.4.050}.

\bibitem{chiaracane2021}
C.~Chiaracane, F.~Pietracaprina, A.~Purkayastha and J.~Goold,
\newblock \emph{Quantum dynamics in the interacting fibonacci chain},
\newblock Phys. Rev. B \textbf{103}, 184205 (2021),
\newblock \doi{10.1103/PhysRevB.103.184205}.

\bibitem{Bonsel2026}
F.~B{\"{o}}nsel, F.~K. Kunst and F.~Roccati,
\newblock \emph{Fibonacci {W}aveguide {Q}uantum {E}lectrodynamics},
\newblock {Quantum} \textbf{10}, 2081 (2026),
\newblock \doi{10.22331/q-2026-04-23-2081}.

\bibitem{sakai2017}
S.~Sakai, N.~Takemori, A.~Koga and R.~Arita,
\newblock \emph{Superconductivity on a quasiperiodic lattice: Extended-to-localized crossover of cooper pairs},
\newblock Phys. Rev. B \textbf{95}, 024509 (2017),
\newblock \doi{10.1103/PhysRevB.95.024509}.

\bibitem{kamiya2018}
K.~Kamiya, T.~Takeuchi, N.~Kabeya, N.~Wada, T.~Ishimasa, A.~Ochiai, K.~Deguchi, K.~Imura and N.~K. Sato,
\newblock \emph{Discovery of superconductivity in quasicrystal},
\newblock Nat. Commun. \textbf{9}(1), 154 (2018),
\newblock \doi{10.1038/s41467-017-02667-x}.

\bibitem{zhang2022}
X.~Zhang and M.~S. Foster,
\newblock \emph{Enhanced amplitude for superconductivity due to spectrum-wide wave function criticality in quasiperiodic and power-law random hopping models},
\newblock Phys. Rev. B \textbf{106}, L180503 (2022),
\newblock \doi{10.1103/PhysRevB.106.L180503}.

\bibitem{wang2024}
Y.~Wang, G.~Rai, C.~Matsumura, A.~Jagannathan and S.~Haas,
\newblock \emph{Superconductivity in the fibonacci chain},
\newblock Phys. Rev. B \textbf{109}, 214507 (2024),
\newblock \doi{10.1103/PhysRevB.109.214507}.

\bibitem{tokumoto2024}
Y.~Tokumoto, K.~Hamano, S.~Nakagawa, Y.~Kamimura, S.~Suzuki, R.~Tamura and K.~Edagawa,
\newblock \emph{Superconductivity in a van der waals layered quasicrystal},
\newblock Nature Communications \textbf{15}(1), 1529 (2024),
\newblock \doi{10.1038/s41467-024-45952-2}.

\bibitem{rai2019}
G.~Rai, S.~Haas and A.~Jagannathan,
\newblock \emph{Proximity effect in a superconductor-quasicrystal hybrid ring},
\newblock Phys. Rev. B \textbf{100}, 165121 (2019),
\newblock \doi{10.1103/PhysRevB.100.165121}.

\bibitem{rai2020}
G.~Rai, S.~Haas and A.~Jagannathan,
\newblock \emph{Superconducting proximity effect and order parameter fluctuations in disordered and quasiperiodic systems},
\newblock Phys. Rev. B \textbf{102}, 134211 (2020),
\newblock \doi{10.1103/PhysRevB.102.134211}.

\bibitem{sandberg2024}
A.~Sandberg, O.~A. Awoga, A.~M. Black-Schaffer and P.~Holmvall,
\newblock \emph{Josephson effect in a fibonacci quasicrystal},
\newblock Phys. Rev. B \textbf{110}, 104513 (2024),
\newblock \doi{10.1103/PhysRevB.110.104513}.

\bibitem{fulga2016}
I.~C. Fulga, D.~I. Pikulin and T.~A. Loring,
\newblock \emph{Aperiodic weak topological superconductors},
\newblock Phys. Rev. Lett. \textbf{116}, 257002 (2016),
\newblock \doi{10.1103/PhysRevLett.116.257002}.

\bibitem{ghadimi2021}
R.~Ghadimi, T.~Sugimoto, K.~Tanaka and T.~Tohyama,
\newblock \emph{Topological superconductivity in quasicrystals},
\newblock Phys. Rev. B \textbf{104}, 144511 (2021),
\newblock \doi{10.1103/PhysRevB.104.144511}.

\bibitem{kobialka2024}
A.~Kobia\l{}ka, O.~A. Awoga, M.~Leijnse, T.~Doma\ifmmode~\acute{n}\else \'{n}\fi{}ski, P.~Holmvall and A.~M. Black-Schaffer,
\newblock \emph{Topological superconductivity in fibonacci quasicrystals},
\newblock Phys. Rev. B \textbf{110}, 134508 (2024),
\newblock \doi{10.1103/PhysRevB.110.134508}.

\bibitem{hori2024}
M.~Hori, T.~Sugimoto, T.~Tohyama and K.~Tanaka,
\newblock \emph{Self-consistent study of topological superconductivity in two-dimensional quasicrystals},
\newblock Phys. Rev. B \textbf{110}, 144512 (2024),
\newblock \doi{10.1103/PhysRevB.110.144512}.

\bibitem{tamura2025}
R.~Tamura, T.~Abe, S.~Yoshida, Y.~Shimozaki, S.~Suzuki, A.~Ishikawa, F.~Labib, M.~Avdeev, K.~Kinjo, K.~Nawa and T.~J. Sato,
\newblock \emph{Observation of antiferromagnetic order in a quasicrystal},
\newblock Nat. Phys.  (2025),
\newblock \doi{10.1038/s41567-025-02858-0}.

\bibitem{Gosh2025}
A.~K. Ghosh, R.~S. Souto, V.~Azimi-Mousolou, A.~M. Black-Schaffer and P.~Holmvall,
\newblock \emph{Quantum state transfer and maximal entanglement between distant qubits using a minimal quasicrystal pump},
\newblock Phys. Rev. B \textbf{112}, 205427 (2025),
\newblock \doi{10.1103/8zys-w2v4}.

\bibitem{lubin2012}
S.~M. Lubin, W.~Zhou, A.~J. Hryn, M.~D. Huntington and T.~W. Odom,
\newblock \emph{High-rotational symmetry lattices fabricated by moiré nanolithography},
\newblock Nano Lett. \textbf{12}(9), 4948 (2012),
\newblock \doi{10.1021/nl302535p}.

\bibitem{joon-ahn2018}
S.~J. Ahn, P.~Moon, T.-H. Kim, H.-W. Kim, H.-C. Shin, E.~H. Kim, H.~W. Cha, S.-J. Kahng, P.~Kim, M.~Koshino, Y.-W. Son, C.-W. Yang \emph{et~al.},
\newblock \emph{Dirac electrons in a dodecagonal graphene quasicrystal},
\newblock Science \textbf{361}(6404), 782 (2018),
\newblock \doi{10.1126/science.aar8412}.

\bibitem{mahmood2021}
R.~Mahmood, A.~V. Ramirez and A.~C. Hillier,
\newblock \emph{Creating two-dimensional quasicrystal, supercell, and moiré lattices with laser interference lithography: Implications for photonic bandgap materials},
\newblock ACS Appl. Nano Mater. \textbf{4}(9), 8851 (2021),
\newblock \doi{10.1021/acsanm.1c00210}.

\bibitem{uri2023}
A.~Uri, S.~C. de~la Barrera, M.~T. Randeria, D.~Rodan-Legrain, T.~Devakul, P.~J.~D. Crowley, N.~Paul, K.~Watanabe, T.~Taniguchi, R.~Lifshitz, L.~Fu, R.~C. Ashoori \emph{et~al.},
\newblock \emph{Superconductivity and strong interactions in a tunable moiré quasicrystal},
\newblock Nature \textbf{620}(7975), 762 (2023),
\newblock \doi{10.1038/s41586-023-06294-z}.

\bibitem{jagannathan21}
A.~Jagannathan,
\newblock \emph{The fibonacci quasicrystal: Case study of hidden dimensions and multifractality},
\newblock Rev. Mod. Phys. \textbf{93}, 045001 (2021),
\newblock \doi{10.1103/RevModPhys.93.045001}.

\bibitem{kohmoto1983}
M.~Kohmoto, L.~P. Kadanoff and C.~Tang,
\newblock \emph{Localization problem in one dimension: Mapping and escape},
\newblock Phys. Rev. Lett. \textbf{50}, 1870 (1983),
\newblock \doi{10.1103/PhysRevLett.50.1870}.

\bibitem{ostlund1983}
S.~Ostlund, R.~Pandit, D.~Rand, H.~J. Schellnhuber and E.~D. Siggia,
\newblock \emph{{One-Dimensional Schr\"odinger Equation with an Almost Periodic Potential}},
\newblock Phys. Rev. Lett. \textbf{50}, 1873 (1983),
\newblock \doi{10.1103/PhysRevLett.50.1873}.

\bibitem{sire89}
{Sire, Clément} and {Mosseri, Rémy},
\newblock \emph{Spectrum of 1d quasicrystals near the periodic chain},
\newblock J. Phys. France \textbf{50}(24), 3447 (1989),
\newblock \doi{10.1051/jphys:0198900500240344700}.

\bibitem{piechon95}
F.~Pi\'echon, M.~Benakli and A.~Jagannathan,
\newblock \emph{Analytical results for scaling properties of the spectrum of the fibonacci chain},
\newblock Phys. Rev. Lett. \textbf{74}, 5248 (1995),
\newblock \doi{10.1103/PhysRevLett.74.5248}.

\bibitem{moustaj23}
A.~Moustaj, M.~Röntgen, C.~V. Morfonios, P.~Schmelcher and C.~Morais~Smith,
\newblock \emph{Spectral properties of two coupled fibonacci chains},
\newblock New Journal of Physics \textbf{25}(9), 093019 (2023),
\newblock \doi{10.1088/1367-2630/acf0e0}.

\bibitem{reisner2023}
M.~Reisner, Y.~Tahmi, F.~Pi\'echon, U.~Kuhl and F.~Mortessagne,
\newblock \emph{Experimental observation of multifractality in fibonacci chains},
\newblock Phys. Rev. B \textbf{108}, 064210 (2023),
\newblock \doi{10.1103/PhysRevB.108.064210}.

\bibitem{kraus12}
Y.~E. Kraus and O.~Zilberberg,
\newblock \emph{Topological equivalence between the fibonacci quasicrystal and the harper model},
\newblock Phys. Rev. Lett. \textbf{109}, 116404 (2012),
\newblock \doi{10.1103/PhysRevLett.109.116404}.

\bibitem{madsen13}
K.~A. Madsen, E.~J. Bergholtz and P.~W. Brouwer,
\newblock \emph{Topological equivalence of crystal and quasicrystal band structures},
\newblock Phys. Rev. B \textbf{88}, 125118 (2013),
\newblock \doi{10.1103/PhysRevB.88.125118}.

\bibitem{jagannathan2025}
A.~Jagannathan,
\newblock \emph{Missing link between the two-dimensional quantum hall problem and one-dimensional quasicrystals},
\newblock Phys. Rev. B \textbf{112}, L100102 (2025),
\newblock \doi{10.1103/stk9-d9vf}.

\bibitem{negro2003}
L.~Dal~Negro, C.~J. Oton, Z.~Gaburro, L.~Pavesi, P.~Johnson, A.~Lagendijk, R.~Righini, M.~Colocci and D.~S. Wiersma,
\newblock \emph{Light transport through the band-edge states of fibonacci quasicrystals},
\newblock Phys. Rev. Lett. \textbf{90}, 055501 (2003),
\newblock \doi{10.1103/PhysRevLett.90.055501}.

\bibitem{steurer2007}
W.~Steurer and D.~Sutter-Widmer,
\newblock \emph{Photonic and phononic quasicrystals},
\newblock J. Phys. D: Appl. Phys. \textbf{40}(13), R229 (2007),
\newblock \doi{10.1088/0022-3727/40/13/r01}.

\bibitem{verbin2013}
M.~Verbin, O.~Zilberberg, Y.~E. Kraus, Y.~Lahini and Y.~Silberberg,
\newblock \emph{Observation of topological phase transitions in photonic quasicrystals},
\newblock Phys. Rev. Lett. \textbf{110}, 076403 (2013),
\newblock \doi{10.1103/PhysRevLett.110.076403}.

\bibitem{tanese2014}
D.~Tanese, E.~Gurevich, F.~Baboux, T.~Jacqmin, A.~Lema\^{\i}tre, E.~Galopin, I.~Sagnes, A.~Amo, J.~Bloch and E.~Akkermans,
\newblock \emph{Fractal energy spectrum of a polariton gas in a fibonacci quasiperiodic potential},
\newblock Phys. Rev. Lett. \textbf{112}, 146404 (2014),
\newblock \doi{10.1103/PhysRevLett.112.146404}.

\bibitem{verbin2015}
M.~Verbin, O.~Zilberberg, Y.~Lahini, Y.~E. Kraus and Y.~Silberberg,
\newblock \emph{Topological pumping over a photonic fibonacci quasicrystal},
\newblock Phys. Rev. B \textbf{91}, 064201 (2015),
\newblock \doi{10.1103/PhysRevB.91.064201}.

\bibitem{baboux2017}
F.~Baboux, E.~Levy, A.~Lema\^{\i}tre, C.~G\'omez, E.~Galopin, L.~Le~Gratiet, I.~Sagnes, A.~Amo, J.~Bloch and E.~Akkermans,
\newblock \emph{Measuring topological invariants from generalized edge states in polaritonic quasicrystals},
\newblock Phys. Rev. B \textbf{95}, 161114(R) (2017),
\newblock \doi{10.1103/PhysRevB.95.161114}.

\bibitem{zilberberg2018}
O.~Zilberberg, S.~Huang, J.~Guglielmon, M.~Wang, K.~P. Chen, Y.~E. Kraus and M.~C. Rechtsman,
\newblock \emph{Photonic topological boundary pumping as a probe of 4d quantum hall physics},
\newblock Nature \textbf{553}(7686), 59 (2018),
\newblock \doi{10.1038/nature25011}.

\bibitem{lisiecki2019}
F.~Lisiecki, J.~Rych\l{}y, P.~Ku\ifmmode~\acute{s}\else \'{s}\fi{}wik, H.~G\l{}owi\ifmmode~\acute{n}\else \'{n}\fi{}ski, J.~W. K\l{}os, F.~Gro\ss{}, N.~Tr\"ager, I.~Bykova, M.~Weigand, M.~Zelent, E.~J. Goering, G.~Sch\"utz \emph{et~al.},
\newblock \emph{Magnons in a quasicrystal: Propagation, extinction, and localization of spin waves in fibonacci structures},
\newblock Phys. Rev. Appl. \textbf{11}, 054061 (2019),
\newblock \doi{10.1103/PhysRevApplied.11.054061}.

\bibitem{goblot2020}
V.~Goblot, A.~Štrkalj, N.~Pernet, J.~L. Lado, C.~Dorow, A.~Lemaître, L.~Le~Gratiet, A.~Harouri, I.~Sagnes, S.~Ravets, A.~Amo, J.~Bloch \emph{et~al.},
\newblock \emph{Emergence of criticality through a cascade of delocalization transitions in quasiperiodic chains},
\newblock Nat. Phys. \textbf{16}(8), 832 (2020),
\newblock \doi{10.1038/s41567-020-0908-7}.

\bibitem{franca2024}
S.~Franca, T.~Seidemann, F.~Hassler, J.~van~den Brink and I.~C. Fulga,
\newblock \emph{Impedance spectroscopy of chiral symmetric topoelectrical circuits},
\newblock Phys. Rev. B \textbf{109}, L241103 (2024),
\newblock \doi{10.1103/PhysRevB.109.L241103}.

\bibitem{Ghosh2026}
A.~K. Ghosh, A.~Chen, A.~E. Hassan, P.~Holmvall, M.~Bourennane and A.~M. Black-Schaffer,
\newblock \emph{Observation of end-to-end pumping in a quasiperiodic fibonacci-type photonic chain} (2026), \eprint{2605.13116}.

\bibitem{yao2019}
H.~Yao, A.~Khoudli, L.~Bresque and L.~Sanchez-Palencia,
\newblock \emph{Critical behavior and fractality in shallow one-dimensional quasiperiodic potentials},
\newblock Phys. Rev. Lett. \textbf{123}, 070405 (2019),
\newblock \doi{10.1103/PhysRevLett.123.070405}.

\bibitem{lahini2009}
Y.~Lahini, R.~Pugatch, F.~Pozzi, M.~Sorel, R.~Morandotti, N.~Davidson and Y.~Silberberg,
\newblock \emph{Observation of a localization transition in quasiperiodic photonic lattices},
\newblock Phys. Rev. Lett. \textbf{103}, 013901 (2009),
\newblock \doi{10.1103/PhysRevLett.103.013901}.

\bibitem{mace2016}
N.~Mac\'e, A.~Jagannathan and F.~Pi\'echon,
\newblock \emph{Fractal dimensions of wave functions and local spectral measures on the fibonacci chain},
\newblock Phys. Rev. B \textbf{93}, 205153 (2016),
\newblock \doi{10.1103/PhysRevB.93.205153}.

\bibitem{ahmed2022}
A.~Ahmed, A.~Ramachandran, I.~M. Khaymovich and A.~Sharma,
\newblock \emph{Flat band based multifractality in the all-band-flat diamond chain},
\newblock Phys. Rev. B \textbf{106}, 205119 (2022),
\newblock \doi{10.1103/PhysRevB.106.205119}.

\bibitem{provost1980}
J.~Provost and G.~Vallee,
\newblock \emph{Riemannian structure on manifolds of quantum states},
\newblock Communications in Mathematical Physics \textbf{76}, 289 (1980).

\bibitem{marzari1997}
N.~Marzari and D.~Vanderbilt,
\newblock \emph{Maximally localized generalized wannier functions for composite energy bands},
\newblock Phys. Rev. B \textbf{56}, 12847 (1997),
\newblock \doi{10.1103/PhysRevB.56.12847}.

\bibitem{marsal2024}
Q.~Marsal and A.~M. Black-Schaffer,
\newblock \emph{Enhanced quantum metric due to vacancies in graphene},
\newblock Phys. Rev. Lett. \textbf{133}, 026002 (2024),
\newblock \doi{10.1103/PhysRevLett.133.026002}.

\bibitem{peotta2015}
S.~Peotta and P.~T{\"o}rm{\"a},
\newblock \emph{Superfluidity in topologically nontrivial flat bands},
\newblock Nat. Commun. \textbf{6}(1), 8944 (2015),
\newblock \doi{https://doi.org/10.1038/ncomms9944}.

\bibitem{tian2023}
H.~Tian, X.~Gao, Y.~Zhang, S.~Che, T.~Xu, P.~Cheung, K.~Watanabe, T.~Taniguchi, M.~Randeria, F.~Zhang \emph{et~al.},
\newblock \emph{Evidence for dirac flat band superconductivity enabled by quantum geometry},
\newblock Nature \textbf{614}(7948), 440 (2023).

\bibitem{bergholtz2013}
E.~J. BERGHOLTZ and Z.~LIU,
\newblock \emph{Topological flat band models and fractional chern insulators},
\newblock International Journal of Modern Physics B \textbf{27}(24), 1330017 (2013),
\newblock \doi{10.1142/S021797921330017X},
\newblock \eprint{https://doi.org/10.1142/S021797921330017X}.

\bibitem{ledwith2020}
P.~J. Ledwith, G.~Tarnopolsky, E.~Khalaf and A.~Vishwanath,
\newblock \emph{Fractional chern insulator states in twisted bilayer graphene: An analytical approach},
\newblock Phys. Rev. Res. \textbf{2}, 023237 (2020),
\newblock \doi{10.1103/PhysRevResearch.2.023237}.

\bibitem{varjas2022}
D.~Varjas, A.~Abouelkomsan, K.~Yang and E.~J. Bergholtz,
\newblock \emph{{Topological lattice models with constant Berry curvature}},
\newblock SciPost Phys. \textbf{12}, 118 (2022),
\newblock \doi{10.21468/SciPostPhys.12.4.118}.

\bibitem{neupert2013}
T.~Neupert, C.~Chamon and C.~Mudry,
\newblock \emph{Measuring the quantum geometry of bloch bands with current noise},
\newblock Phys. Rev. B \textbf{87}, 245103 (2013),
\newblock \doi{10.1103/PhysRevB.87.245103}.

\bibitem{ozawa2019}
T.~Ozawa and N.~Goldman,
\newblock \emph{Probing localization and quantum geometry by spectroscopy},
\newblock Phys. Rev. Res. \textbf{1}, 032019 (2019),
\newblock \doi{10.1103/PhysRevResearch.1.032019}.

\bibitem{kang2025}
M.~Kang, S.~Kim, Y.~Qian, P.~M. Neves, L.~Ye, J.~Jung, D.~Puntel, F.~Mazzola, S.~Fang, C.~Jozwiak, A.~Bostwick, E.~Rotenberg \emph{et~al.},
\newblock \emph{Measurements of the quantum geometric tensor in solids},
\newblock Nature Physics \textbf{21}, 110 (2025),
\newblock \doi{10.1038/s41567-024-02678-8}.

\bibitem{verma2026}
N.~Verma, P.~J. Moll, T.~Holder and R.~Queiroz,
\newblock \emph{Quantum geometry and the hidden scales in materials},
\newblock Nature Reviews Physics pp. 1--14 (2026).

\bibitem{berry1984}
M.~V. Berry,
\newblock \emph{Quantal phase factors accompanying adiabatic changes},
\newblock Proceedings of the Royal Society of London. A. Mathematical and Physical Sciences \textbf{392}(1802), 45 (1984).

\bibitem{bianco2011}
R.~Bianco and R.~Resta,
\newblock \emph{Mapping topological order in coordinate space},
\newblock Phys. Rev. B \textbf{84}, 241106 (2011),
\newblock \doi{10.1103/PhysRevB.84.241106}.

\bibitem{sticlet2013}
D.~Sticlet and F.~Pi\'echon,
\newblock \emph{Distant-neighbor hopping in graphene and haldane models},
\newblock Phys. Rev. B \textbf{87}, 115402 (2013),
\newblock \doi{10.1103/PhysRevB.87.115402}.

\bibitem{yu2025}
J.~Yu, B.~A. Bernevig, R.~Queiroz, E.~Rossi, P.~T{\"o}rm{\"a} and B.-J. Yang,
\newblock \emph{Quantum geometry in quantum materials},
\newblock npj Quantum Materials \textbf{10}(1), 101 (2025).

\bibitem{mace17}
N.~Mac\'e, A.~Jagannathan and F.~Piéchon,
\newblock \emph{Gap structure of 1d cut and project hamiltonians},
\newblock Journal of Physics: Conference Series \textbf{809}(1), 012023 (2017),
\newblock \doi{10.1088/1742-6596/809/1/012023}.

\bibitem{morfonios2014}
C.~Morfonios, P.~Schmelcher, P.~Kalozoumis and F.~Diakonos,
\newblock \emph{Local symmetry dynamics in one-dimensional aperiodic lattices: a numerical study},
\newblock Nonlinear Dynamics \textbf{78}, 71 (2014).

\bibitem{wang2025}
J.~Wang, Y.~Chen and H.~Huang,
\newblock \emph{Quantum metric enhancement and hierarchical scaling in one-dimensional quasiperiodic systems}  (2025),
\newblock \eprint{2507.04213}.

\bibitem{varma2016}
V.~K. Varma, S.~Pilati and V.~E. Kravtsov,
\newblock \emph{Conduction in quasiperiodic and quasirandom lattices: Fibonacci, riemann, and anderson models},
\newblock Phys. Rev. B \textbf{94}, 214204 (2016),
\newblock \doi{10.1103/PhysRevB.94.214204}.

\bibitem{carrasco2025}
H.~Roche~Carrasco, J.~Schirmann, A.~Mordret and A.~G. Grushin,
\newblock \emph{Family of aperiodic tilings with tunable quantum geometric tensor},
\newblock Phys. Rev. Lett. \textbf{135}, 236603 (2025),
\newblock \doi{10.1103/dzqm-9kwj}.

\bibitem{katz1986}
{Katz, A.} and {Duneau, M.},
\newblock \emph{Quasiperiodic patterns and icosahedral symmetry},
\newblock J. Phys. France \textbf{47}(2), 181 (1986),
\newblock \doi{10.1051/jphys:01986004702018100}.

\bibitem{duneau1985}
M.~Duneau and A.~Katz,
\newblock \emph{Quasiperiodic patterns},
\newblock Phys. Rev. Lett. \textbf{54}, 2688 (1985),
\newblock \doi{10.1103/PhysRevLett.54.2688}.

\bibitem{elser1986}
V.~Elser,
\newblock \emph{{The diffraction pattern of projected structures}},
\newblock Acta Crystallographica Section A \textbf{42}(1), 36 (1986),
\newblock \doi{10.1107/S0108767386099932}.

\bibitem{kraus2016}
Y.~E. Kraus and O.~Zilberberg,
\newblock \emph{Quasiperiodicity and topology transcend dimensions},
\newblock Nature Physics \textbf{12}(7), 624,
\newblock \doi{10.1038/nphys3784}.

\bibitem{Wang2023}
N.~Wang, D.~Kaplan, Z.~Zhang, T.~Holder, N.~Cao, A.~Wang, X.~Zhou, F.~Zhou, Z.~Jiang, C.~Zhang, S.~Ru, H.~Cai \emph{et~al.},
\newblock \emph{Quantum-metric-induced nonlinear transport in a topological antiferromagnet},
\newblock Nature \textbf{621}(7979), 487 (2023),
\newblock \doi{https://doi.org/10.1038/s41586-023-06363-3}.

\bibitem{li2021}
Z.~Li, S.~Zhang, T.~Tohyama, X.~Song, Y.~Gu, T.~Iitaka, H.~Su and H.~Zeng,
\newblock \emph{Optical detection of quantum geometric tensor in intrinsic semiconductors},
\newblock Science China Physics, Mechanics \& Astronomy \textbf{64}, 107211 (2021),
\newblock \doi{10.1007/s11433-021-1750-2}.

\bibitem{julku2016}
A.~Julku, S.~Peotta, T.~I. Vanhala, D.-H. Kim and P.~T\"orm\"a,
\newblock \emph{Geometric origin of superfluidity in the lieb-lattice flat band},
\newblock Phys. Rev. Lett. \textbf{117}, 045303 (2016),
\newblock \doi{10.1103/PhysRevLett.117.045303}.

\bibitem{you88}
J.~Q. You and T.~B. Hu,
\newblock \emph{A global cut-and-project method to construct generalized fibonacci lattices and quasilattices},
\newblock physica status solidi (b) \textbf{147}(2), 471 (1988),
\newblock \doi{https://doi.org/10.1002/pssb.2221470203},
\newblock \eprint{https://onlinelibrary.wiley.com/doi/pdf/10.1002/pssb.2221470203}.

\bibitem{rontgen19}
M.~R\"ontgen, C.~V. Morfonios, R.~Wang, L.~Dal~Negro and P.~Schmelcher,
\newblock \emph{Local symmetry theory of resonator structures for the real-space control of edge states in binary aperiodic chains},
\newblock Phys. Rev. B \textbf{99}, 214201 (2019),
\newblock \doi{10.1103/PhysRevB.99.214201}.

\bibitem{wehling2010}
T.~O. Wehling, S.~Yuan, A.~I. Lichtenstein, A.~K. Geim and M.~I. Katsnelson,
\newblock \emph{Resonant scattering by realistic impurities in graphene},
\newblock Phys. Rev. Lett. \textbf{105}, 056802 (2010),
\newblock \doi{10.1103/PhysRevLett.105.056802}.

\bibitem{moustaj2021}
A.~Moustaj, S.~Kempkes and C.~M. Smith,
\newblock \emph{Effects of disorder in the fibonacci quasicrystal},
\newblock Phys. Rev. B \textbf{104}, 144201 (2021),
\newblock \doi{10.1103/PhysRevB.104.144201}.

\bibitem{ssh1979}
W.~P. Su, J.~R. Schrieffer and A.~J. Heeger,
\newblock \emph{Solitons in polyacetylene},
\newblock Phys. Rev. Lett. \textbf{42}, 1698 (1979),
\newblock \doi{10.1103/PhysRevLett.42.1698}.

\end{thebibliography}


\end{document}